\newcommand \Mpc {h^{-1}{\rm Mpc}}
\newcommand \kpc {h^{-1}{\rm kpc}}

\newcommand \arcs{\hbox{$^{\prime\prime}$}}

\newcommand \kms {{\rm km~s}^{-1}}

\newcommand \beqn {\begin{equation}}
\newcommand \eeqn {\end{equation}}

\newcommand \ncztot {15767 }

\newcommand \ha {H$\alpha$~}
\newcommand \ewha {EW[H$\alpha$]}
\newcommand \ewhd {EW[H$\delta$]}
\newcommand{\I}{{\scshape i\@}}
\newcommand{\II}{{\scshape ii\@}}
\newcommand{\III}{{\scshape iii\@}}

\documentclass[12pt,preprint]{aastex}
\usepackage{emulateapj5}
\usepackage{onecolfloat}   

\begin{document}

\twocolumn[

\title{CAIRNS: The Cluster And Infall Region Nearby Survey 
III. Environmental Dependence of H$\alpha$ Properties of Galaxies}
\shorttitle{CAIRNS III. H$\alpha$ Properties and Environment}
\shortauthors{Rines et al.}

\author{Kenneth Rines\altaffilmark{1}, Margaret J. Geller\altaffilmark{2},
Michael J. Kurtz\altaffilmark{2}, \& Antonaldo Diaferio\altaffilmark{3}}
\email{krines@astro.yale.edu}

\altaffiltext{1}{Yale Center for Astronomy and Astrophysics, Yale University, P.O. Box 208121, New Haven CT 06520-8121; krines@astro.yale.edu}
\altaffiltext{2}{Smithsonian Astrophysical Observatory; mgeller, mkurtz@cfa.harvard.edu}
\altaffiltext{3}{Universit\`a degli Studi di Torino,
Dipartimento di Fisica Generale ``Amedeo Avogadro'', Torino, Italy; diaferio@ph.unito.it}

\begin{abstract}

We investigate the environmental dependence of star formation in
cluster virial regions and infall regions as part of CAIRNS (Cluster
And Infall Region Nearby Survey), a large spectroscopic survey of the
infall regions surrounding nine nearby rich clusters of galaxies.  We
use complete, homogeneous spectroscopic surveys of $K_s$ limited
samples in eight of the CAIRNS clusters.  Our long-slit spectroscopy
yields estimates of star formation rates in environments from cluster
cores to the general large-scale structure.  Galaxies in infall
regions probe whether processes affecting star formation are effective
over scales larger than cluster virial regions.  The fraction of
galaxies with current star formation in their inner disks as traced by
\ha emission increases with distance from the cluster and converges to
the ``field'' value only at 2-3 virial radii, in agreement with
other investigations.  However, among galaxies with significant
current star formation (\ewha$\geq$2\AA), there is no difference in
the {\it distribution} of \ewha~ inside and outside the virial radius.
This surprising result, first seen by Carter et al., suggests that (1)
star formation is truncated on either very short timescales or only at
moderate and high redshifts or (2) that projection effects contaminate
the measurement.  We quantify the possible impact of mechanisms which
only affect the outer parts of galaxies and thus might not be detected
in this survey or any fiber-based survey.  The number density profiles
of star-forming and non-star-forming galaxies indicate that, among
galaxies projected inside the virial radius, at least half of the
former and 20\% of the latter are ``infall interlopers,'' galaxies in
the infall region but outside the virial region.  We show that the
kinematics of star-forming galaxies in the infall region closely match
those of absorption-dominated galaxies.  This result shows that the
star forming galaxies in the infall regions are not interlopers from
the field and excludes one model of the backsplash scenario of galaxy
transformation.  Finally, we quantify systematic uncertainties in
estimating the global star formation in galaxies from their inner
disks.

\end{abstract}


]

\section{Introduction}

One of the most striking features of galaxy clusters at the present
epoch is their deficit of late-type galaxies relative to the overall
galaxy population in magnitude limited samples.  This feature was
noted by \citet{1931ApJ....74...43H}, and eventually quantified by
\citet{dressler80}.  Similar trends are seen in galaxy groups
\citep{1984ApJ...281...95P}, suggesting that groups may play an
important role in establishing the morphology-density relation.  This
morphological segregation is reflected in segregation of spectral
types, i.e., galaxies in dense environments are much less likely than
those in less dense environments to show evidence of ongoing star
formation in their spectra \citep[e.g.,][and references
therein]{1989ApJ...347..152S,bcarter}.  A third method of galaxy
classification is to separate them by their broadband colors.
Morphology, color, and spectral type are well correlated, but there
are galaxies such as ``passive spirals'' that would be classified
differently in these different schemes
\citep{1998ApJ...497L..75K,goto03}.  Here, we focus on spectral
information, in particular \ha emission, which traces current star
formation.

Recently, several investigators have found that these population
differences are not restricted to cluster virial regions.  Galaxy
properties seem to converge to those of field galaxies only at 2-3
virial radii \citep{1996ApJ...471..694A,balogh97,diaferio01,ellingson01,lewis02,gomez03,balogh03,gray04,tanaka04}.
The two leading explanations for the presence of these galaxies are
preprocessing by infalling groups (and/or filaments) and
``backsplash''.  In the backsplash model, some galaxies at large radii
have passed through the virial region of the cluster and thus may have
undergone environmental transformation within the virial radius.  

These differences may result from the types of galaxies that form in
different environments or from environmental processes (or more
likely, some mixture of the two).  Truly environmental mechanisms are
strongly implicated by the observation that cluster galaxies have
different star formation properties than field galaxies with similar
stellar mass, morphology, and mean stellar age \citep{christlein05}.
Several physical mechanisms can alter star formation in galaxies as
they enter dense environments \citep[see][for a clear summary of these
mechanisms]{treu03}.  Infall regions (where galaxies are infalling
onto the main cluster but have not yet reached equilibrium) provide a
unique probe of the environmental scale which determines galaxy
properties: infall regions are overdense averaged over scales of
5-10$\Mpc$ (because they are near a cluster) but they contain a wide
range of local densities measured on scales of roughly 1$\Mpc$ (e.g.,
groups and filaments).

Ongoing massive star formation produces \ha emission from H\II~
regions.  The strength of this emission line is a good measure of the
current star formation rate.  \ha is relatively insensitive to
metallicity, unlike some star formation tracers including [O\II]
\citep{kewley04}.  The existence of galaxies with little or no star
formation but late-type morphologies suggests that star formation is
more sensitive to environment than morphology
\citep{1976ApJ...206..883V,1998ApJ...497L..75K,goto03,vogt04b,kodama04}.

Here, we study the \ha properties of a large, homogeneous
spectroscopic sample of galaxies in the virial regions and infall
regions of eight nearby clusters from CAIRNS \citep[The Cluster And
Infall Region Nearby Survey, see ][hereafter Paper I]{cairnsi}.  This
survey is the largest sample of long-slit spectra in cluster infall
regions to date.  Even with long-slit spectra, the spectral properties
are weighted towards those of the inner disks (radii of $\sim$3 kpc).
Because CAIRNS probes rich clusters, we probe densities up to an order
of magnitude larger than studies based on 2dF and SDSS (which at
present contain relatively few rich, nearby clusters).  The
spectroscopic catalogs are complete in the near-infrared to absolute
magnitude $M_{K_s}=-22.7 + 5 \mbox{log} h$ using photometry from
2MASS, the Two-Micron All-Sky Survey \citep{twomass}.  This limit is
approximately one magnitude fainter than $M_{K_s}^*$, the
characteristic magnitude of the luminosity function averaged over all
environments \citep{twomdflfn,cairnsii}.  Our selection of galaxies in
the near-infrared leads to a sample that is much closer to a sample
selected by stellar mass than samples selected at optical wavelengths.
Because of the shape of the galaxy luminosity function, a substantial
fraction of any magnitude or luminosity limited sample consists of
galaxies close to the faint limit.  In optically selected samples, the
stellar masses of star forming galaxies near the survey limit are
smaller than the stellar masses of galaxies near the survey limit
without star formation.  Thus, near infrared selection significantly
reduces this potential systematic bias.

There are two related but separable issues when studying star
formation rates in cluster galaxies.  The first is the fraction of the
total galaxy population with significant current star formation.
Bright galaxies in the cores of clusters show little evidence of
recent star formation; we probe the radial extent of this effect.  The
second issue is the distribution of star formation rates among
galaxies with current star formation.  Separating these issues should
help clarify the physical mechanisms responsible for the
transformation \citep{balogh03}.  For instance, mechanisms which
truncate star formation on short timescales (like ram pressure
stripping) would likely produce a relatively sharp population gradient
but leave the distribution of SFRs unchanged.  Mechanisms which
operate over several Gigayears would likely produce milder population
gradients but with a larger change in the SFR distribution (i.e.,
dense regions would contain more galaxies with small but non-zero
SFRs).

We describe the observations in $\S \ref{obs}$.  We discuss the
distribution of galaxies with and without \ha emission in $\S
\ref{distrib}$.  In $\S \ref{distribha}$, we discuss the distribution
of equivalent widths in different environments.  We discuss our
results in $\S \ref{discuss}$ and conclude in $\S \ref{conclusions}$.
We assume a cosmology of $H_0 = 100~h~\kms, \Omega_m = 0.3, 
\Omega_\Lambda = 0.7$ throughout.

\section{Observations \label{obs}}

\subsection{The CAIRNS Cluster Sample}

We selected the CAIRNS parent sample from all nearby
($cz<15,000~\kms$), Abell richness class $R\geq1$ \citep[][]{aco1989},
X-ray luminous ($L_X>2.5 \times 10^{43} h^{-2}$erg s$^{-1}$) galaxy
clusters with declination $\delta>-15^\circ$. Using X-ray data from
the X-ray Brightest Abell Clusters catalog \citep{xbacs}, the parent
cluster sample contains 14 systems.  We selected a representative
sample of 8 of these 14 clusters (Table \ref{sample}).  The cluster
properties listed in Table \ref{sample} are from Paper I.  The 6
clusters meeting the selection criteria but not targeted in CAIRNS
are: A193, A426, A2063, A2107, A2147, and A2657.  The 8 CAIRNS
clusters span a variety of morphologies, from isolated clusters (A496,
A2199) to major mergers (A168, A1367).  Note that we omit A576 from
the present sample because it lacks uniform, homogeneous spectroscopy
($\S 2.3$).

\begin{table*}[th] \footnotesize
\begin{center}
\caption{\label{sample} \sc CAIRNS Parent Population}
\begin{tabular}{lcccccccc}
\tableline
\tableline
\tablewidth{0pt}
Cluster &\multicolumn{2}{c}{X-ray Coordinates} & $cz_\odot$ & $\sigma
 _p$ & $L_X$ & $T_X$ & Richness  \\ 
 & RA (J2000) & DEC (J2000) &  $\kms$ & $\kms$ & $10^{43}
 h^{-2}$~ergs~s$^{-1}$& keV  \\ 
\tableline
A119 & 00 56 12.9 & -01 14 06 & 13268 &698$^{+36}_{-31}$ & 8.1 & 5.1 & 1  \\
A168 & 01 15 08.8 & +00 21 14 & 13395 &579$^{+36}_{-30}$ & 2.7  & 2.6 & 2  \\
A496 & 04 33 35.2 & -13 14 45 & 9900 &721$^{+35}_{-30}$ & 8.9  & 4.7 & 1  \\
A539 & 05 16 32.1 & +06 26 31 & 8717 &734$^{+53}_{-44}$ & 2.7  & 3.0 & 1 \\ 
A576\tablenotemark{a} & 07 21 31.6 & +55 45 50 & 11510 &1009$^{+41}_{-36}$ & 3.5 & 3.7 & 1  \\
A1367 & 11 44 36.2 & +19 46 19 & 6495 &782$^{+56}_{-46}$ & 4.1  & 3.5 & 2  \\ 
Coma & 12 59 31.9 & +27 54 10 & 6973 &1042$^{+33}_{-30}$ & 18.0  & 8.0 & 2  \\
A2199 & 16 28 39.5 & +39 33 00 & 9101 &796$^{+38}_{-33}$ & 9.1  & 4.7 & 2  \\
\tableline
A194 & 01 25 50.4 & -01 21 54 & 5341 &495$^{+41}_{-33}$ & 0.4  & 2.6 & 0  \\
\tableline
\tablenotetext{a}{Omitted from the present sample because it lacks uniform spectroscopy.}
\end{tabular}
\end{center}
\end{table*}

The redshift limit is set by the small aperture of the 1.5-m
Tillinghast telescope used for the vast majority of our spectroscopic
observations. The richness minimum guarantees that the systems contain
sufficiently large numbers of galaxies to sample the velocity
distribution.  The X-ray luminosity minimum guarantees that the
systems are real clusters and not superpositions of galaxy groups
\citep[cf. the discussion of A2197 in][]{rines01a, rines02}.  
Three additional clusters with smaller X-ray luminosities (A147, A194
and A2197) serendipitously lie in the survey regions of A168 and
A2199.  A147 and A2197 lie at nearly identical redshifts to A168 and
A2199; their dynamics are probably dominated by the more massive
cluster \citep{rines02}.  A194, however, is cleanly separated from
A168 and we therefore analyze it as a ninth system.  The inclusion of
A194 extends the parameter space covered by the CAIRNS sample.  The
X-ray temperature of A194 listed in \citet{xbacs} is an extrapolation
of the $L_X - T_X$ relation; in Table \ref{sample} we therefore list the
direct temperature estimate of \citet{1998PASJ...50..187F} from {\em
ASCA} data. \citet{1998PASJ...50..187F} list X-ray temperatures for 6
of the 9 CAIRNS clusters which agree with those listed in
\citet{xbacs}. 

In Paper I, we applied a hierarchical clustering analysis
\citep[described in][]{diaferio1999} to the redshift catalogs to
determine the central coordinates and redshift of the largest system
of galaxies in each cluster.  We adopt these hierarchical centers as
the cluster centers.  We then used the caustic technique to compute
mass profiles for the clusters to very large radii.  We define the
cluster virial region as the volume inside $r_{200}$ ($r_\delta$ is
the radius within which the enclosed mass density is $\delta \rho_c$,
where $\rho_c$ is the critical density).  The infall region is the
volume between $r_{200}$ and $r_{3.5}$, which corresponds to the
turnaround radius $r_t$ of a spherical overdensity \citep{rg89}.
Empirically, $r_t\approx 5r_{200}$.  Note that $R_\delta$ refers to
$r_\delta$ in projection; it has the same numerical value and is used
only to emphasize projected quantities.

\subsection{2MASS Photometry}

2MASS is an all-sky survey with uniform, complete photometry
\citep{twomasscalib} in three infrared bands (J, H, and K$_s$, a
modified version of the K filter truncated at longer wavelengths). We
use photometry from the final extended source catalog
\citep[XSC,][]{twomassxsc}.  The 2MASS XSC computes magnitudes in the
$K_s$-band using several different methods, including aperture
magnitudes (using a circular aperture with radius 7\arcs), isophotal
magnitudes which include light within the elliptical isophote
corresponding to $\mu_{K_s}$=20 mag/arcsec$^2$, Kron magnitudes, and
extrapolated ``total'' magnitudes \citep{twomassxsc}.  The sky
coverage of the catalog is complete except for small regions around
bright stars.

The 2MASS isophotal magnitudes omit $\sim$15\% of the total flux of
individual galaxies (K01).  C01 compare 2MASS photometry from the
Second Incremental Data Release (2IDR) with deeper infrared photometry
from \citet{loveday00}.  They find that Kron magnitudes are slightly
fainter than the total magnitudes in deeper surveys \citep[see
also][]{andreon02} and that 2MASS extrapolated total magnitudes are
slightly brighter than Kron magnitudes (for the deeper survey these
are close to total magnitudes) from the deeper survey.
2MASS is a relatively shallow survey and thus likely misses many low
surface brightness galaxies \citep{andreon02,bell03}.  In this work we
focus on bright galaxies (which typically have high surface
brightness); thus, this bias is less important here than for estimates
of the global luminosity density or stellar mass density.  Except
where noted, we define ``bright galaxies'' as those with
$M_{K_s}=-22.7 + 5 \mbox{log} h$.

We use the $K_s$-band survey extrapolated ``total'' magnitudes.
Galactic extinction is minor in the near-infrared.  We correct for
Galactic extinction by using the value in the center of the cluster
which we estimate from the dust maps of \citet{sfd98}.  We make K
corrections and evolutionary corrections of $<$0.15 magnitudes based
on \citet{pogg97}.  Because these corrections are small and not
strongly dependent on the galaxy model at the redshifts of the CAIRNS
clusters, we apply a uniform correction for all galaxies in a given
cluster interpolated from the model Elliptical SED with solar
metallicity and a star-formation e-folding time of 1 Gyr.

\subsection{Spectroscopy \label{spectra}}

We have collected \ncztot redshifts within a radius of $\sim$10$\Mpc$
of the 9 clusters in the CAIRNS sample (Paper I) and an additional 515
redshifts to obtain complete near-infrared selected samples
\citep{cairnsii}.  The spectra were obtained with the FAST
spectrograph \citep{fast} on the 1.5-m Tillinghast telescope of the
Fred Lawrence Whipple Observatory (FLWO).  FAST is a high throughput,
long slit spectrograph with a thinned, backside illuminated,
antireflection coated CCD detector.  The slit length is 180\arcs; our
observations used a slit width of 3\arcs ~and a 300 lines mm$^{-1}$
grating.  The slit orientation is fixed at 90$^\circ$ (E-W) for nearly
all observations to maximize observing efficiency.  This setup yields
spectral resolution of 6-8\AA~and covers the wavelength range
3600-7200\AA.  We obtain redshifts by cross-correlation with spectral
templates of emission-dominated and absorption-dominated galaxy
spectra created from FAST observations
\citep{km98}.  The typical systematic uncertainty in the redshifts is
30$~\kms$; the statistical uncertainty is usually comparable.

The target catalogs were constructed first with photographic plates
(see details in Paper I) and later with 2MASS \citep{cairnsii}.  The
redshift catalogs are essentially complete to at least $M_{K_s}=-22.7
+ 5 \mbox{log} h$ (our definition of ``bright galaxies'') with some
sampling of fainter galaxies from the optical catalogs.  This limit is
roughly 1 magnitude fainter than $M^*_{K_s}$ for field galaxies
\citep{twomdflfn} and contains $\sim$60\% of the total cluster light
(in galaxies) assuming that the cluster and field LFs are identical.
The completeness limits in absolute magnitudes of the clusters lie in
the range -22.70$\leq M_{K_s,lim}-5\mbox{log}h\leq$-21.42; we use
these slightly deeper limits to reduce statistical uncertainties when
studying individual cluster properties with small numbers of ``bright
galaxies''.  We drop A576 from the sample because most of the spectra
in the cluster core were obtained with different instrumentation
\citep{mohr96}.  Rather than correct for differences in
instrumentation for this one cluster, we restrict our study to
clusters with substantially complete FAST spectroscopy.  Table
\ref{samplen} lists the number of bright galaxies in the virial region
of each cluster ($R_p\leq R_{200}$), in the infall region
(1$<R_p/R_{200}\leq$5), and the sum of the two.  The radius $r_{200}$
is the radius within which the enclosed matter density is 200 times
the critical value ($R_{200}$ is this radius for projected radii).  In
a critical density universe, $r_{200}$ is about equal to the virial
radius; in a $\Lambda$CDM universe, the virial radius is approximately
$r_{100}\approx1.4r_{200}$.  Table \ref{samplen} shows that we have
FAST spectra for 92.4\% of bright galaxies with $R_p/R_{200}\leq$5.
Note that A119 and A168 have a projected separation of $10.7~\Mpc
\approx 9.7 R_{200}$.   Thus, the outskirts of A119 and A168
overlap.

\begin{table*}[th] \footnotesize
\begin{center}
\caption{\label{samplen} \sc CAIRNS Virial and Infall Region Membership}
\begin{tabular}{lccc}
\tableline
\tableline
\tablewidth{0pt}
Cluster & $N_{200}$ & $N_{inf}$ & $N_{tot}$ \\
\tableline
A119 & 52(40) & 71(67) & 123(107) \\
A168 & 38(25) & 79(74) & 117(99) \\
A496 & 40(38) & 40(37) & 80(77) \\
A539 & 32(21) & 26(22) & 58(43) \\
A1367 & 44(44) & 58(57) & 102(101) \\
Coma & 97(97) & 115(115) & 212(212) \\
A2199 & 55(50) & 159(145) & 214(195) \\
A194 & 20(20) & 20(20) & 40(40) \\
\tableline
All & 378(335) & 568(539) & 946(874) \\
Field & & & 656 \\
\tableline

\tablenotetext{a}{Numbers in parentheses are the number of galaxies with FAST spectra.}
\end{tabular}
\end{center}
\end{table*}

An important difference between the FAST spectra collected for CAIRNS
and those collected for other, larger redshift surveys
\citep{2df,sdss} is that the S/N is similar for bright and faint galaxies.
CAIRNS suffers no incompleteness due to fiber
placement constraints.
Another difference is that the long-slit FAST spectra sample light
from larger fractions of the galaxies than fiber spectra.  Thus, the
effects of aperture bias \citep[e.g.,][]{apbias,kewley05} on spectral
classification are greatly reduced.  \citet{bcarter} show that a
spectroscopic survey of field galaxies obtained with identical
instrumentation in a similar redshift range contains no significant
aperture bias.  

However, there is an important caveat about aperture bias.  Many Virgo
spirals \citep{koopmann04b} as well as some spirals in other nearby
clusters \citep{vogt04b} show evidence of truncated \ha disks.
Because these truncated disks are rare in samples of field
galaxies, estimates of aperture bias based on field samples
\citep{bcarter,kewley05} may not apply to cluster samples.  We discuss
this issue further in $\S \ref{discuss}$.  The CAIRNS spectra should
present an unbiased picture of the star formation properties of the
inner parts of galaxies (radii $\lesssim$3 kpc), but future studies
(with either narrowband imaging and/or integrated spectra) are
required to see whether these relations accurately represent global
star formation properties.

Our spectroscopic catalog contains many foreground and background
galaxies.  We can use these spectra to investigate possible systematic
differences between galaxies in cluster, infall, and field
environments.  Note, however, that our catalog of field galaxy spectra
is not complete.  In particular, we do not have spectra for
(preferentially bright) galaxies with previously measured redshifts
which we classify as foreground or background galaxies.  We use field
galaxy spectra only in the redshift interval 2000-12000 $\kms$ to
minimize the effects of aperture corrections and to avoid Virgocentric
infall (Table \ref{samplen}).  Curiously, the distribution of absolute
magnitudes $M_{K_s}$ for our field galaxies with FAST spectra is
extremely similar to that of cluster/infall members (a K-S test fails
to differentiate them at the 90\% confidence level despite having
samples of over 600 galaxies), while both differ significantly from
the distribution of galaxies without FAST spectra.  This happy
circumstance means that our field sample is well matched to our
cluster and infall region sample.

\subsection{Spectroscopic Indices and Types}

We measure spectroscopic properties of the galaxies in several
emission lines.  We extract 1D spectra from the 2D spectra selecting
apertures along the slit which maximize the S/N.  The resulting 1D
spectra thus oversample the inner few kpcs of the galaxies relative to
integrated spectra.  We calculate equivalent widths in the lines from
the 1D spectra using the bandpasses for lines and continua listed in
Table \ref{bandpasses}.  These bandpasses are the ones used by
\citet{bcarter} to study galaxies in the 15R survey except for
H$\beta$, for which we modify the continuum region so that it excludes
the [O\III]$\lambda$4949 line.  Because we focus on emission lines, we
adopt the convention that emission lines have positive equivalent
width.

\begin{table*}[th] \caption{Line Index Definitions\label{bandpasses}}
\begin{tabular}{rlccc}\tableline\tableline
\multicolumn{2}{c}{Index} & Blue Continuum & Line Region & Red Continuum \\
\ & \  & (\AA) & (\AA) & (\AA) \\ \tableline
{[}O\II{]}  & 3727.3\AA & 3653.0---3716.3 & 3716.3---3738.3 & 3738.3---3803.0 \\
H$\beta$  & 4861.3\AA & 4771.3---4831.3 & 4841.3---4881.3 & 4891.3---4951.3 \\
{[}O\III{]} & 5006.8\AA & 4891---4945     & 4995---5019     & 5021---5087 \\
{[}N\II{]}  & 6548.1\AA & 6505---6535     & 6538.1---6558.1 & 6597---6627 \\
H$\alpha$ & 6562.8\AA & 6505---6535     & 6554.5---6574.5 & 6597---6627 \\
{[}N\II{]}  & 6583.4\AA & 6505---6535     & 6573.4---6593.4 & 6597---6627 \\ \tableline
\end{tabular}
\end{table*}

We divide the sample into two main subsamples of galaxies with and
without emission lines (we refer to the latter sample as absorption
galaxies).  We define the emission line sample as galaxies with
\ewha$\geq$2.0\AA.  We do not correct for stellar absorption, which is
typically $\approx$1.0\AA~\citep{balogh03}.  We also do not correct
\ewha ~for absorption by dust because we assume that the extinction to
continuum and to H\II ~regions are identical.  If this assumption is
incorrect \citep{calzetti01}, then Balmer decrements are necessary to
obtain accurate SFR estimates.  However, the environmental trends we
investigate here should be robust to this effect unless the relative
extinctions of continuum and H\II ~regions depend on environment.
The median statistical uncertainty in \ewha ~is $\approx$0.3\AA; the
spectral classification of emission/non-emission lines is robust (less
than 3\% of galaxies have \ewha~within 1$\sigma$ of 2.0\AA).  Our
cutoff for emission line galaxies is slightly more generous than that
of \citet{balogh03}, who uses \ewha$\geq$4\AA~(after adding 1.0\AA~to
the EW to correct roughly for \ha absorption).  We do not convert
\ewha ~to fluxes because few of the spectra have significant
detections of H$\beta$ required to estimate the extinction correction.
The single criterion \ewha$\geq$2\AA~leads to some contamination of
the absorption galaxy subsample by galaxies with obscured star
formation and/or obscured AGN.  For this reason and because the
signal-to-noise ratios of the spectra are not uniform across the
sample, this definition may misclassify galaxies with weak emission
lines in their spectra as absorption line galaxies. The classification
is therefore a separation of gas-rich (including AGN) and gas-poor
systems, with some contamination of the gas-poor subsample.

We directly compare several FAST spectra in the fields of A119 and
A168 to SDSS spectra and find that there is very good agreement in
spectral classification (the presence or absence of \ha emission).
There is substantial, but not surprising, scatter in the values of
\ewha, probably due to the inhomogeneous nature of \ha ~emission
(e.g., the inclusion or exclusion of a single HII region can
significantly affect the inferred \ewha).  A full comparison of FAST
and SDSS spectra is beyond the scope of this work.

We further divide the subsample of galaxies with emission lines into
AGN-dominated galaxies and star forming galaxies using the line
diagnostics of \citet{kewley01}.  The lack of accurate H$\beta$ data
(due to stellar absorption and weaker emission lines) prevents this
classification from being very robust; thus we inspect all the
emission-line spectra interactively with the IRAF routine {\em splot}
to include AGN with no detected H$\beta$ or where broad lines affect
the line indices.  These diagnostics only classify emission as
dominated by AGN if it is impossible to reproduce the emission lines
with H\II ~region photoionization models.  The diagnostics of
\citet{kewley01} classify some galaxies as star forming that would be
classified by \citet{vo87} as AGN.  The AGN fraction is small among
bright $K_s$-selected galaxies; within 5$R_{200}$, the AGN fraction is
2.8$\pm$0.6\%.  This fraction is smaller than the $\sim$20\% in the
studies of \citet{bcarter} and \citet{kauffmann03b}, mostly due to the
different diagnostics used.  The AGN fraction is sufficiently small
that we ignore the AGN/H\II ~classification in the next section where
we discuss the environmental dependence of the fraction of galaxies
with emission lines.

\section{Distribution of Spectroscopic Types \label{distrib}}

\subsection{Radial Distribution of Spectroscopic Types \label{radial}}

\begin{figure*}[tb]
\centerline{\epsfxsize=6in\epsffile{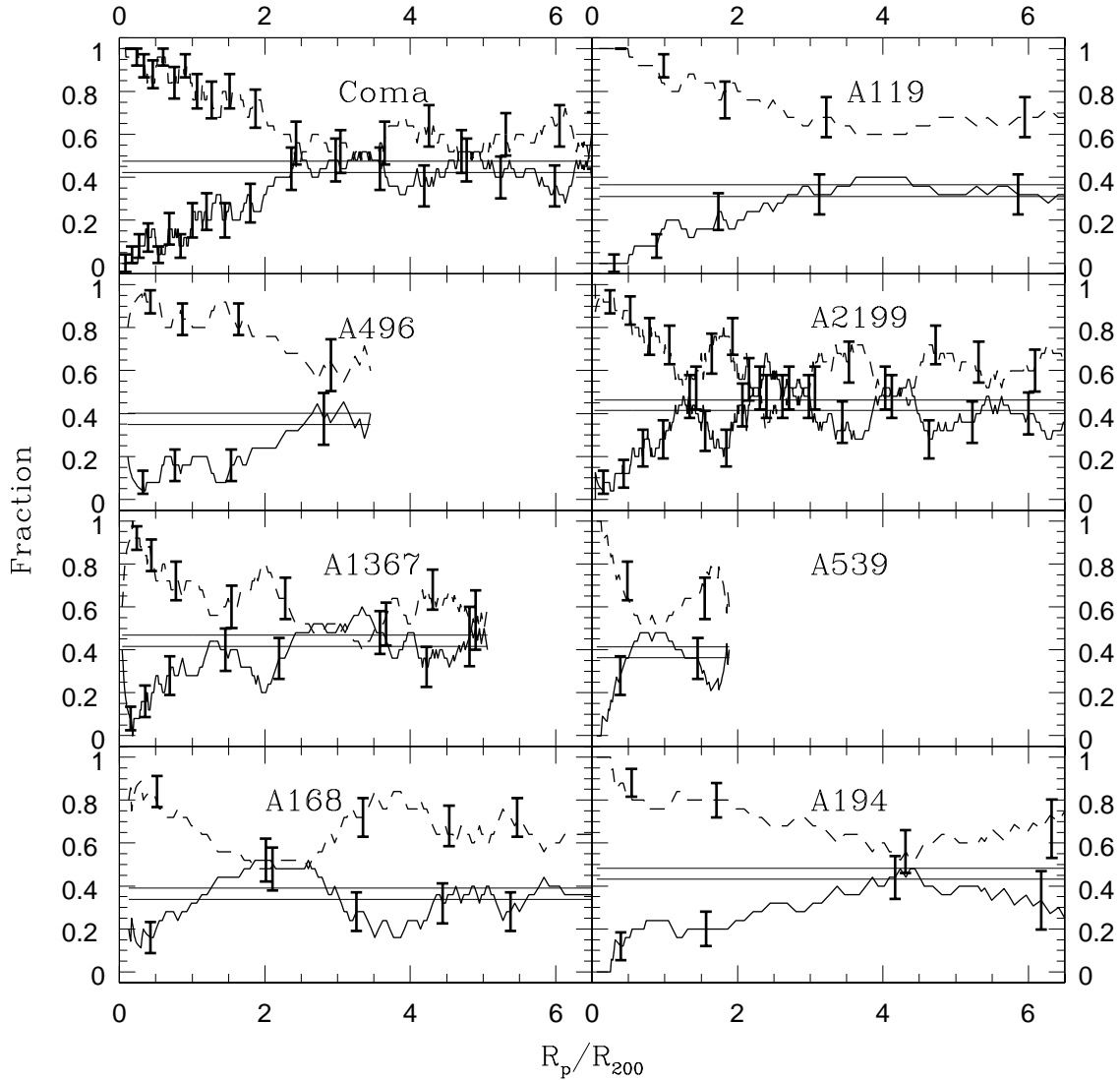}}
\caption{\label{allemfrac} 
Fraction of bright 
 galaxies with (solid lines) and without (dashed lines) \ha emission
as a function of distance from the cluster center in units of
$R_{200}$.  The lines show the fractions in moving averages of 25
galaxies and the errorbars are placed at independent bins and indicate
Poisson uncertainties.  The samples for each cluster are selected by
absolute magnitude ($M_{K_s}\leq M_{K_s,lim}$) with a slightly
different limit for each cluster.  Horizontal lines in each panel show
the 1-$\sigma$ range of the fraction of emission line galaxies in
field samples to the same limiting luminosity.}
\end{figure*}

Figure \ref{allemfrac} shows the fraction of galaxies (with FAST
spectra) in the emission/absorption subsample as a function of
clustrocentric radius.  The solid lines in each panel show the
expected fraction of emission line galaxies in a field sample with the
same limiting absolute magnitude as the cluster in that panel.  We
estimate this fraction from the field galaxy spectra in our catalog in
the redshift interval 2000-12000 $\kms$.  The fraction of bright
($M_{K_s}\leq -22.7 + 5 \mbox{log} h$) galaxies with emission lines is
similar to the fraction in the complete magnitude-limited survey of
\citet{bcarter}.  The limiting absolute magnitudes of the clusters lie
in the range -22.70$\leq M_{K_s,lim}-5\mbox{log}h\leq$-21.42; the
fractions of emission line galaxies in similarly selected ``field''
samples lie in the range 0.27$\leq f_{ELG}\leq$0.46, similar to the
fraction in the complete magnitude-limited survey of \citet{bcarter}.
Figure \ref{allemfrac} and a similar figure in $\S \ref{fracdensity}$
are the only results shown for a cluster-dependent $M_{K_s,lim}$ rather
than for only ``bright galaxies'' (defined by $M_{K_s}\leq -22.7 +
5\mbox{log} h$).  The emission fraction profiles to a fixed luminosity
cutoff are very similar but have poorer statistics, especially for the
poor cluster A194.  The outskirts of A119 and A168 overlap ($\S
\ref{spectra}$); Figure \ref{allemfrac} shows that the emission-line
fraction at the midpoint is roughly equal to the field value.

\begin{figure}[tb]
\plotone{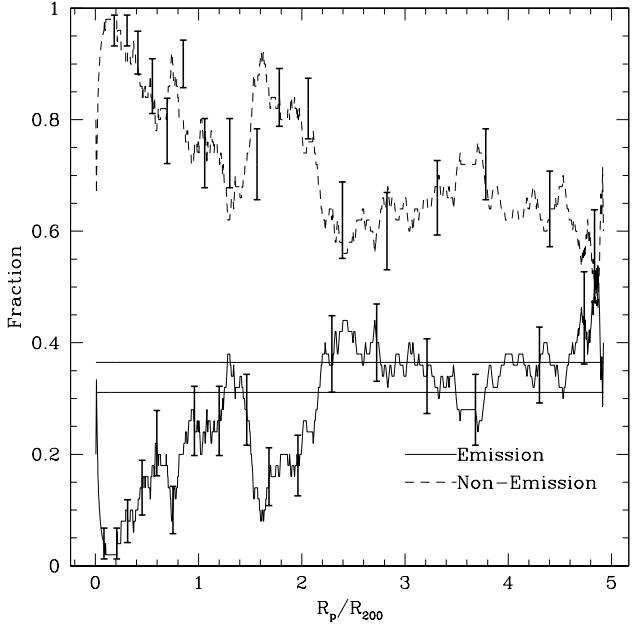} 
\caption{\label{emfracall} Fraction of bright 
 galaxies ($M_{K_s}\leq -22.7$) with (solid lines) and without (dashed
lines) \ha emission as a function of distance from the composite
cluster center in units of $R_{200}$.  The lines show the fractions in
moving averages of 50 galaxies and the errorbars are placed at
independent bins and indicate Poisson uncertainties.  Horizontal lines
show the 1-$\sigma$ range of the fraction of emission line galaxies in
field samples to the same limiting luminosity.}
\end{figure}

We confirm that the fraction of galaxies with current star
formation does not reach the field value until 2-3 $R_{200}$
($R_{200}$ roughly equals the virial radius) from the centers of
clusters
\citep{1996ApJ...471..694A,balogh97,diaferio01,ellingson01,lewis02,gomez03,balogh03}.
Figure \ref{allemfrac} demonstrates that this effect is observable in
individual clusters, although there appear to be cluster-to-cluster
variations in the exact shape of this profile.  As in Paper I,
clusters are ordered left to right and top to bottom in order of
decreasing X-ray temperature, a good proxy for virial mass. The
gradients appear smoothest for the hottest clusters, although the
coolest cluster (A194) also shows a smooth gradient.  Figure
\ref{emfracall} shows the emission fraction in the composite CAIRNS
cluster; the emission fraction converges to the field value at
$\approx$2$R_{200}$.  Because the cluster masses are well determined
(the statistical uncertainties in the corrected virial masses are
$\lesssim 15\%$; see Paper I), the uncertainties in $r_{200}\propto
M^{1/3}$ should not affect the cluster-to-cluster comparison or the
construction of the composite cluster.  Assuming that the population
variation is caused by environment, this result can be interpreted in
two ways.  Either (1) infalling galaxies are ``preprocessed'' at high
densities in filaments and groups in the infall regions, or (2) a
significant number of galaxies projected at large radii have already
passed near the cluster center and are observed on ``first outfall''
or second infall.  The ``preprocessing'' interpretation implicates
relatively local processes such as galaxy-galaxy interactions; the
latter ``backsplash'' scenario implicates processes such as ram
pressure stripping that are strongest near cluster centers.

The infall region of A2199 offers an interesting discriminant between
the two scenarios because it contains several X-ray emitting groups
\citep{rines01b,rines02}.  Under the preprocessing scenario we would
expect that the fraction of galaxies with emission lines would not
converge to the field value until well outside $r_{200}$.  The
fraction of emission-line galaxies in A2199 quickly rises to the field
value just outside $R_{200}$, but the fraction then {\em decreases} at
the radius of the infalling X-ray groups A2197W and A2197E (Figure
\ref{allemfrac}).  This result supports the preprocessing scenario.

Unlike the other clusters, the emission fraction in A168 shows only a
very weak radial trend.  Part of this difference results from the
unrelaxed nature of A168 (Paper I); the X-ray and optical centers are
offset by 160 $\kpc$.  When we plot the emission fraction versus
distance from the X-ray center, the trend is closer to those of the
other CAIRNS clusters.  Although these cluster-to-cluster variations
are interesting, we caution the reader that the appearance of clusters
can vary dramatically depending on projection effects
\citep[e.g.,][]{diaferio1999}; thus one must be cautious not to
overinterpret these variations.

\begin{figure}[tb]
\plotone{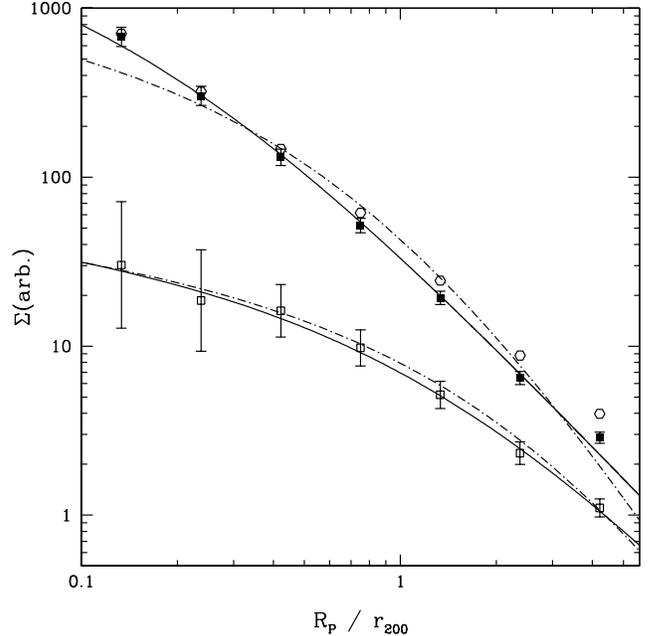} 
\caption{\label{dhist} Surface number density profiles of bright 
 galaxies with (open squares) and without (filled squares) \ha emission
in units of $R_{200}$ (open hexagons show the sum).  The solid (dash-dotted) lines show the best-fit NFW and (Hernquist) profiles. }
\end{figure}

Figure \ref{dhist} shows the number density profiles of emission and
absorption galaxies.  Because the caustics which we use to define
membership extend to a different radius $R_{max}$ for each cluster
(Paper I), the profiles are incompletely sampled (and therefore
underestimated) outside $\sim 2 R_{200}$.  We do not weight by cluster
richness; each galaxy has an equal weight.  Both profiles are fit well
by NFW profiles \citep{nfw97}, with scale radii of
$\approx0.23r_{200}$ for the absorption galaxies and $\approx
1.25r_{200}$ for the emission-line galaxies.  The implied NFW
concentration parameter $c=r_{200}/r_s$ is therefore 4.3 (3.3-5.9 at
68\% confidence) and 0.8 (0.4-1.7 at 68\% confidence) for absorption
and emission line galaxies respectively, and $c\approx 3.3$ for all
galaxies combined.  This last number agrees with the results of a
separate study which uses statistical background subtraction to
determine membership \citep{lin04}.  For comparison, the caustic mass
profiles yield estimates of $c=5-17$ for the CAIRNS clusters.  Cluster
galaxies are therefore less concentrated than dark matter in the
cluster, although the concentration of absorption-line galaxies is
similar to the smallest values of $c$ of the mass profiles.

\citet{1997ApJ...476L...7C} fit \citet{hernquist1990} profiles
to the number density profiles of blue and red samples from the CNOC1
survey.  Using Hernquist profiles, the scale radius of the
absorption-line galaxy sample in CAIRNS ($1.13\pm0.18$) exceeds the
scale radius ($0.56\pm0.10$) of red galaxies in the CNOC1 survey
\citet{1997ApJ...476L...7C}, and the scale radius of the emission line
galaxies ($3.2^{+2.1}_{-1.2}$) exceeds that of the blue galaxies in
CNOC1 ($1.82\pm0.27$).  This comparison suggests that emission-line
galaxies are a subsample of blue galaxies with a more extended
distribution and that absorption-line galaxies are a combination of
red and blue galaxies.  Indeed, 30\% of CNOC1 cluster galaxies are
classified as blue; only 10\% of CAIRNS cluster galaxies inside
$R_{200}$ are emission-line galaxies.  The different scale radii may
reflect the fact that CNOC1 includes few galaxies outside $R_{200}$.

\begin{figure}[tb]
\plotone{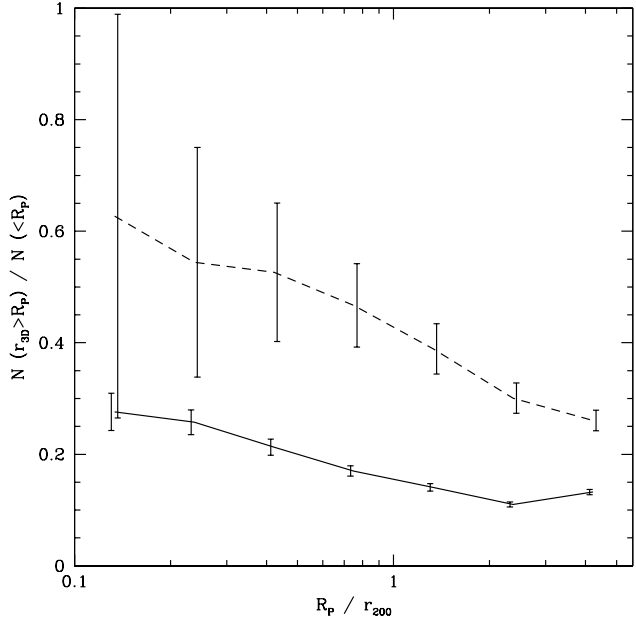} 
\caption{\label{interlopers} Fraction of galaxies within projected 
radius $R_p$ that are ``infall interlopers'' (galaxies within the
cluster virial or infall region but with physical radii $r_{3D}>R_p$)
versus $R_p$, computes assuming NFW number density profiles for both
populations (Figure \ref{dhist}).  The errorbars show the Poissonian
uncertainties on the number density profiles and do not include any
uncertainty in the model. Dashed and solid lines represent emission
and absorption galaxies respectively.}
\end{figure}

Assuming that the NFW profiles accurately represent the true profiles,
we can estimate the number of ``infall interlopers,'' galaxies with
$R_p\leq R_{200}$ but $r_{3D}>r_{200}$.  At least 20\% of absorption
line galaxies and 50\% of emission line galaxies are infall
interlopers.  Figure \ref{interlopers} shows the fraction of galaxies
projected within $R_p$ with physical radii $r_{3D}>R_p$.  The fraction
of infall interlopers increases dramatically as the projected radius
$R_p$ decreases.  This result is consistent with the simulations of
\citet{diaferio01}, who show the fraction of galaxies with
$r_{3D}>r_{200}$ as a function of $R_p$ (see their Figure 13).  The
actual fraction of infall interlopers probably varies significantly
from cluster to cluster due to the non-uniform distribution of
galaxies in the infall region (i.e., groups and filaments).  This
result validates an assumption often used in cluster studies, namely,
that star-forming/blue/spiral galaxies are not good tracers of the
mass distribution in the virial region
\citep[e.g.,][]{1997ApJ...476L...7C}.  It is interesting to note that
even among absorption-line galaxies, the fraction of infall
interlopers is 20\%.  Because the density of the infall region
increases with redshift \citep{ellingson01}, the fraction of infall
interlopers probably increases with redshift.

Projection effects have important implications for the analysis of
cluster galaxies.  The strong radial dependence of the fraction of
galaxies with emission lines suggests that many and perhaps all
emission-line galaxies projected inside $r_{200}$ actually lie at
larger radii.  These galaxies may lie in the infall region or they may
be interlopers from the field.  The ``infall interlopers'' alone
comprise at least 50\% of emission-line galaxies with $R_p\leq
r_{200}$.

Among bright galaxies, the fraction of galaxies without emission lines
is still quite large well outside $R_{200}$.  This result shows that
some early-type galaxies projected inside $R_{200}$ lie at greater
radii and thus that their kinematics might not represent a relaxed
population.  Restricting virial mass estimates to galaxies without
emission lines is effective at removing the emission-dominated
``infall interlopers'' (galaxies inside $r_t$ but outside $r_{200}$)
but does not remove the infall interlopers without emission lines.

\subsection{Kinematic Distribution of Spectroscopic Types}

\begin{figure}[tb]
\plotone{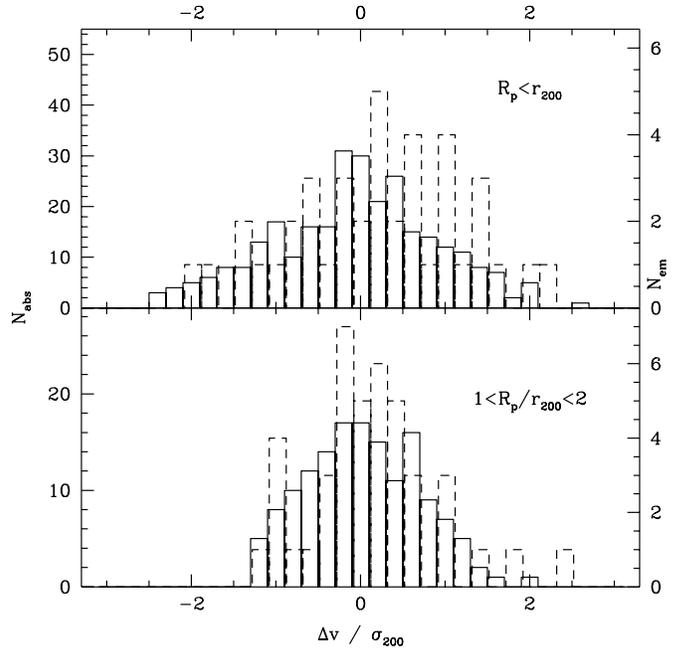} 
\caption{\label{vhist01} (Top panel) Velocity distribution of galaxies 
projected inside $R_{200}$ with (dashed line) and without (solid line)
\ha emission.  (Bottom panel) Same as top panel for galaxies projected 
in the interval $1-2 R_{200}$.  Backsplash galaxies should be peaked
at zero velocity. }
\end{figure}

Figure \ref{vhist01} shows the velocity distribution of emission- and
absorption-line galaxies projected inside $R_{200}$.  We find no
evidence for kinematic segregation; a two-sample K-S test indicates a
28\% probability that two samples drawn from the same parent
distribution would show a larger difference.  We do find a larger
velocity dispersion for the emission-line galaxies with an F test, but
this test assumes that the parent distributions are Gaussian.  Bright
emission-line galaxies are quite rare in clusters; even after stacking
the nine CAIRNS clusters, there are 42 bright emission-line galaxies
inside $R_{200}$, compared to 293 bright absorption-line galaxies (43
bright galaxies have no FAST spectra).

\begin{figure}[tb]
\plotone{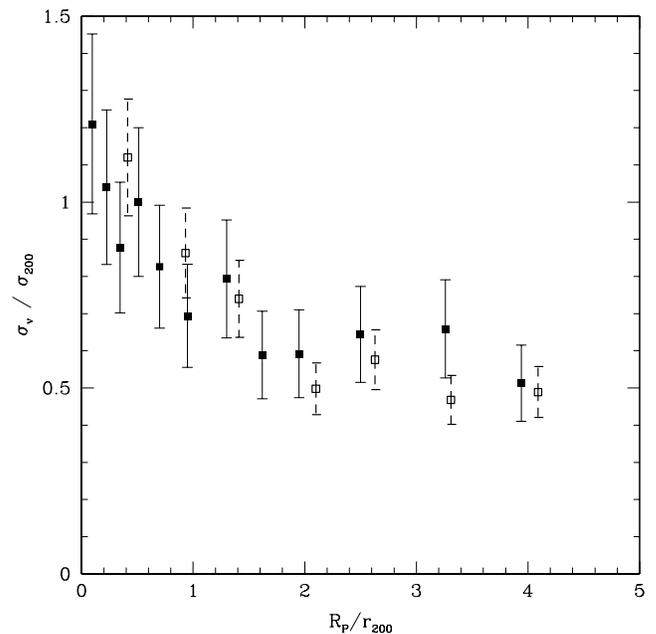} 
\caption{\label{sigmar} Velocity dispersion profiles of  bright galaxies 
with (dashed line) and without (solid line)
\ha emission.   }
\end{figure}

\begin{figure*}[tb]
\centerline{\epsfxsize=6in\epsffile{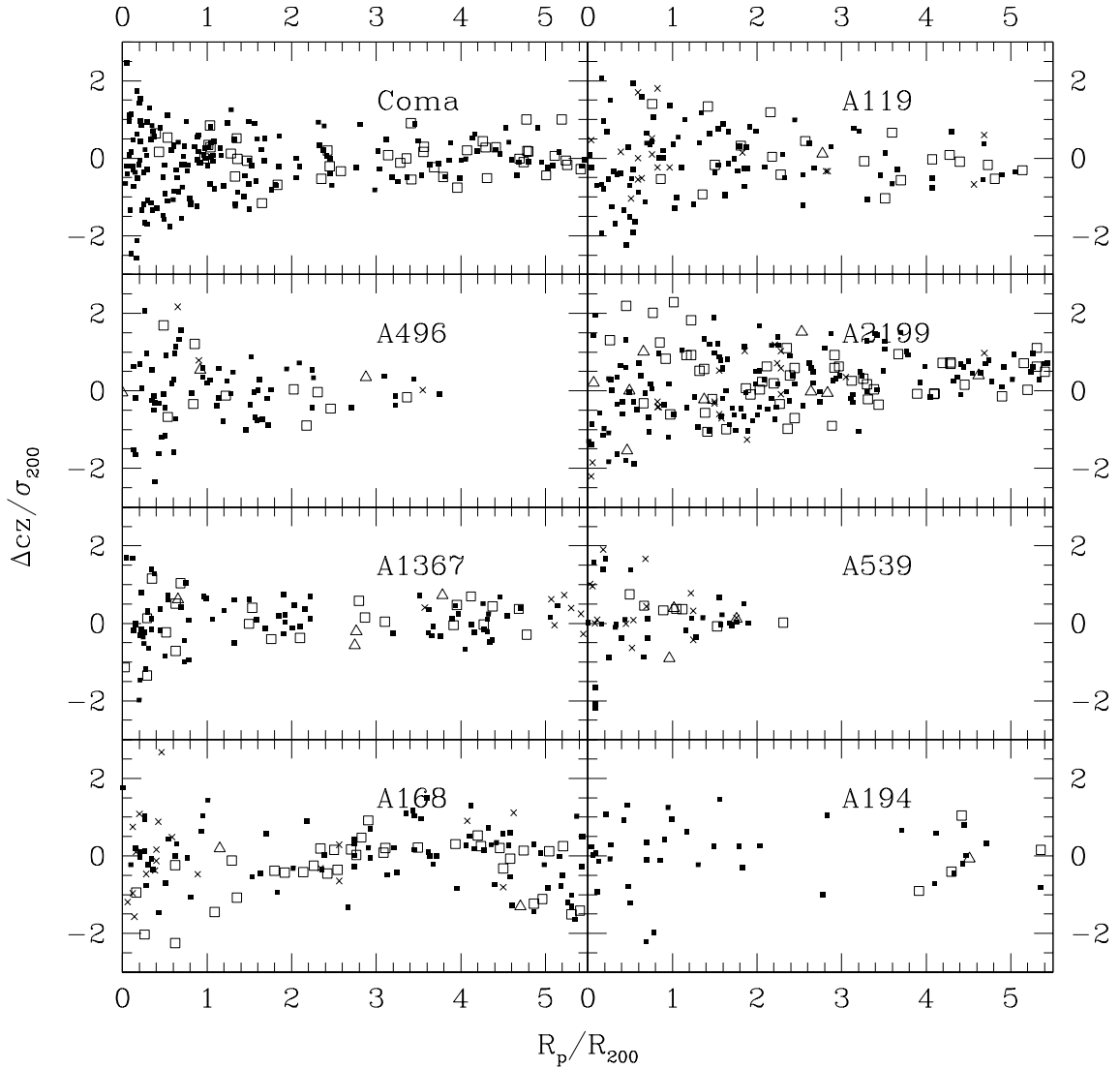}}
\caption{\label{causspec} Velocity versus projected radius for bright galaxies 
with (open squares and triangles) and without (dots) \ha emission (crosses are galaxies without FAST spectra).  Squares and triangles indicate H\II and AGN type emission respectively.}
\end{figure*}

Figure \ref{sigmar} shows the velocity dispersion profiles (VDPs) of
emission and absorption line galaxies.  The profiles are quite
similar.  We obtain similar results when including all galaxies
brighter than $M_{K_s}=-21.0$ (an incomplete sample), suggesting that
the similarity of the VDPs extends to fainter magnitudes.  There is
also no significant difference between the velocity distributions in
the interval 1$<R_p/R_{200}\leq$2 (bottom panel of Figure
\ref{vhist01}).  Figure \ref{causspec} shows redshift versus projected
radius for the CAIRNS member galaxies with different symbols for
emission and absorption line galaxies.

These results contradict conventional wisdom about the velocity
distribution of early and late-type galaxies.  Early studies of
kinematic differences in Virgo \citep{huchra85,binggeli87} show
different velocity dispersions for early and late-type galaxies, but
Virgo is a very complex system and neither investigation uses a
nonparametric K-S test.  Similar differences exist in A576 for
emission- and absorption-dominated galaxies \citep{mohr96}, and for
red and blue galaxies projected in the virial regions of clusters
\citep{kg82,cd96,1997ApJ...476L...7C}.  \citet{mohr96} found that the
velocity dispersions differ according to an F test, but ``a K-S test
fails to distinguish between the two velocity distributions.''  In the
CNOC1 clusters, red and blue galaxies have significantly different
VDPs, but Jeans' analysis yields consistent cluster masses for both
red and blue populations \citep{1997ApJ...476L...7C}.  The CNOC1
results therefore suggest that the blue galaxy population is in
dynamical equilibrium with the same underlying mass profile as the red
galaxies.  A galaxy population on first infall is not expected to be
in equilibrium with the underlying mass profile and would not
necessarily satisfy the Jeans' equations.  The emission-line galaxies
in CAIRNS have the same velocity distribution but a different spatial
distribution than the absorption-dominated galaxies; the CAIRNS
emission-line galaxies are therefore not in dynamical equilibrium with
the mass profile traced by the absorption-dominated galaxies.

To test the robustness of previous results, we revisit the analysis of
\citet{cd96}, who found a difference between red and blue galaxies at
the 99.2\% confidence level with a K-S test in the Coma cluster.  They
use the photometric catalog of \citet{gmp83} based on photographic
plates to identify galaxies within and blueward of a color-magnitude
relation (see their Figure 12).  We reproduce their color cut using an
updated redshift catalog from NED and Paper I.  Using a similar
membership criterion ($4000\leq cz \leq 10000~\kms$), we find that the
difference is only significant at 97.5\%; the significance drops to
90.0\% with the redshift cut $4000\leq cz \leq 9000~\kms$.  Their
result is based only on galaxies classified in the larger of two
subclusters defined by the KMM algorithm; we perform a similar cut by
selecting only galaxies within 30$'$ of the center of Coma, but obtain
nearly identical results (97.6\% versus 97.5\%).  We therefore
conclude that the kinematic difference between blue and red galaxies
claimed by \citet{cd96} is not robust.

The inclusion of galaxies at the edges of the distributions risks
including interlopers from the field; because field galaxies have a
much larger fraction of blue galaxies, these interlopers are likely to
be blue galaxies.  Inclusion of these field interlopers artificially
increases the velocity dispersion of blue galaxies.  Our membership
selection from the caustics is generally more conservative than
techniques such as 3-$\sigma$ clipping, which are used for sparser
galaxy samples.  If there are real differences in the kinematics of
red and blue galaxies, these should be present in the main part of the
distribution and not just in the high-velocity tails.  Thus, stricter
membership criteria should not mask real kinematic differences.
However, looser criteria may include interlopers and artificially
increase the velocity dispersion of blue galaxies relative to red
galaxies.

\citet{biviano02} performed a more robust investigation of luminosity 
and morphological segregation in the ENACS clusters.  They find
significant differences in the (R,v) distributions of three classes of
galaxy types (E+E/S0+S0, early-type spirals, and late-type spirals +
emission line galaxies).  However, they find no significant difference
in the velocity distributions of E+E/S0+S0 and early-type spirals in
the radial range $0.25\leq R_p/r_{200}\leq 0.75$ or in the three types
in the radial range $0.75\leq R_p/r_{200}\leq 1.5$.  Possible
explanations for the differing results include the classification used
(morphology versus emission lines), the sampling of fainter galaxies
in ENACS (this is particularly important for later morphological types
which are rare in CAIRNS), and differences in membership assignment.

The issue of kinematic segregation among galaxy types is very subtle.
The results depend significantly on membership classification, survey
completeness and uniformity, survey depth, morphological versus
spectroscopic classification, and the definitions of cluster centers.
CAIRNS uses relatively strict membership criteria which may classify
galaxies at the edges of the velocity distribution as non-members.
This selection tends to decrease the difference between the two
populations.


\subsection{A Test of the ``Backsplash'' Scenario}

Recent simulations demonstrate that a significant number of galaxies
observed outside cluster virial radii are ``backsplash'' galaxies
\citep{bnm2000,2004A&A...414..445M}, galaxies that have passed through
the virial region.  H\I ~observations of galaxies in Virgo suggest
that galaxies at radii as large as $2R_{200}$ have undergone ram
pressure stripping
\citep{2002AJ....124.2440S,2002ApJ...580..164S,vogt04b}, a process
which requires that the galaxies have encountered dense intracluster
gas present only in the core of Virgo.  \citet{2004A&A...418..393S}
show that these galaxies can be backsplash galaxies.

\begin{figure}[tb]
\plotone{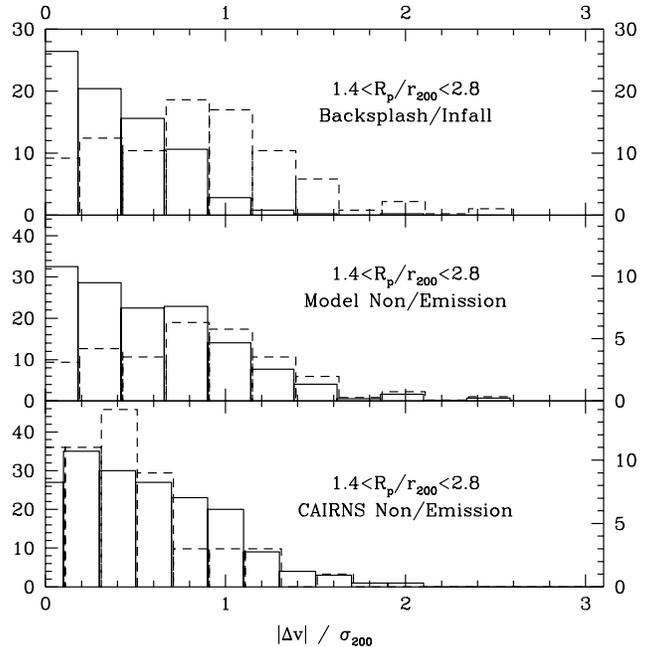} 
\caption{\label{backsplash} (Top panel) Velocity distribution of 
model cluster galaxies projected in the interval $1.4-2.8 R_{200}$
from the backsplash and infalling populations (solid and dashed lines
respectively).  (Middle panel) Velocity distribution of absorption and
emission galaxies (solid and dashed lines respectively) under a toy
model where all backsplash galaxies are absorption-dominated and
infalling galaxies are split evenly between those with and without \ha
emission.  (Bottom panel) Observed distribution of emission/absorption
galaxies (dashed/solid lines) in the CAIRNS clusters.}
\end{figure}

Recent simulations by \citet{gill04} show that the velocity
distribution of backsplash galaxies is much more centrally peaked than
that of infalling galaxies (see their Figure 8; they define backsplash
galaxies as those which have passed within $1.4 r_{200}$).  If all
galaxies in the interval $1.4<R_p/R_{200}<2.8$ are on first infall,
then the observed absolute velocity ($|cz-cz_{cluster}|$) distribution
peaks at $\sim$400$\kms$.  If, instead, galaxies in this interval are
a mixture of backsplash and infalling galaxies, then the observed
absolute velocity distribution is peaked at zero velocity.  Thus,
\citet{gill04} suggest that the observed shape of the total velocity
histogram in the interval $1.4<R_p/R_{200}<2.8$ can test whether
backsplash galaxies are present in this interval or whether all
galaxies in this interval are on first infall (see their Figure 8,
reproduced here as the top panel of Figure \ref{backsplash}).  We
compare the total velocity distribution of CAIRNS galaxies in this
radial interval with a mock total velocity distribution of backsplash
plus infalling galaxies from Figure 8 of \citet{gill04}.  A K-S test
fails to distinguish between the two total velocity distributions at
the 95\% confidence level.  A K-S test clearly distinguishes between
the CAIRNS total velocity distribution and both the velocity
distributions of infalling and backsplash galaxies (greater than
99.9\% confidence).  Thus, our observations of the total velocity
distribution show clear evidence for the existence of a mixed
backsplash and infalling population.  This result excludes extreme
scenarios where all backsplash galaxies are either completely
disrupted or sufficiently stripped of stars to lie below our minimum
luminosity.

While the total velocity histogram provides evidence of the existence
of a backsplash population, it does not contain information on the
{\em properties} of backsplash and infalling galaxies.  \citet{gill04}
show that the velocities of backsplash galaxies are much more strongly
peaked around the cluster redshift than the velocities of galaxies
infalling for the first time.  When projected along an observer's
line-of-sight, these differences are reduced but still significant.
Thus, if the properties of a galaxy reflect whether it is backsplash
or on first infall, then different galaxy distributions will have
different velocity distributions.  For instance, if close passage to a
cluster center always truncates star formation, then a galaxy with
emission lines must be on first infall and not a backsplash galaxy.
We therefore propose a toy model of the backsplash scenario to convert
the simulation results into observable quantities: we assume that
environmental truncation of star formation in galaxies is achieved
only by passage through the virial region and that the truncation
mechanism is perfectly efficient. 


In this toy model, the velocity distribution of emission-line (and
therefore all infalling) galaxies should be skewed to larger projected
velocities relative to absorption-line galaxies (which contain a
mixture of backsplash and infalling galaxies) (middle panel of Figure
\ref{backsplash}).  From the CAIRNS field sample, 66$\pm$5\%
of bright field galaxies lack emission lines, so we assume that 66\%
of bright galaxies on first infall will lack star formation due to
non-environmental effects.  The difference between the velocity distributions
of the two populations should be statistically significant even if
66\% of the infalling galaxies are absorption-line galaxies (a K-S
test shows a difference at $\gtrsim$99\% confidence for a sample the
size of CAIRNS; middle panel of Figure \ref{backsplash}).  Note that
\citet{gill04} convolve the projected velocities with a `typical'
observational error of 100$\kms$; because the uncertainties in our
redshifts are typically 30$\kms$, their Figure 8 understates the
expected difference between the populations for the CAIRNS sample.  

We now turn to the observations.  The bottom panel of Figure
\ref{backsplash} shows the velocity distribution of emission and
absorption-line galaxies from CAIRNS in the interval
$1.4<R_p/R_{200}<2.8$.  While our toy model predicts significant
differences, the observed velocity distributions of emission line and
absorption line galaxies are indistinguishable (consistent at the 89\%
level with a K-S test; see the bottom panel of Figure
\ref{backsplash}).

The similarity of the velocity distributions of star-forming and
passive galaxies in the interval $1.4<R_p/R_{200}<2.8$ shows that this
toy model of the backsplash scenario disagrees with the data.
Assuming that the kinematic differences between backsplash and
infalling galaxies predicted by the simulations are correct, then some
emission-line galaxies must be backsplash galaxies (otherwise the
velocity histogram of the emission-line galaxies in the bottom panel
of Figure \ref{backsplash} would not be centrally peaked).  That is,
the observed velocity distribution of emission-line galaxies suggests
that at least some of them have passed through the virial region
($1.4r_{200}$) of the cluster.  Thus, whether a galaxy is a backsplash
galaxy or on first infall is not the primary determinant of the
presence or absence of ongoing star formation (in the inner disk).
The above toy model of the backsplash scenario could probably be
modified to fit the data, for example by restricting the definition of
backsplash galaxies to those that have passed within 0.5$r_{200}$
(rather than within $1.4r_{200}$) of the central cluster or by
allowing the mechanism affecting backsplash galaxies to be less than
100\% efficient.  It is also possible that such a passage may strip
the halo gas reservoir of a galaxy that would fuel future
star-formation but would not strip the existing molecular gas,
allowing continued star formation for $\sim$1 Gyr \citep[][]{larson80}
or only strip the outer disk, leading to truncated spirals
\citep{koopmann04b,vogt04b} which may not be detected in our data.  We
exclude only the extreme model described above.

The conclusion that a pure backsplash model does not describe the data
is consistent with the conclusion of previous investigations that the
fraction of galaxies with current star formation shows the same
dependence on local density both inside and outside clusters
\citep[][see $\S \ref{fracdensity}$]{lewis02,gomez03,balogh03}.  That
is, both analyses suggest that mechanisms correlated with local
density are more important than proximity to a cluster in determining
a galaxy's star formation rate.

Another important implication of the kinematic similarity between
star-forming and passive galaxies is that the star-forming galaxies
are not all interloping field galaxies; field interlopers would not
show a velocity peak at the cluster redshift.  Most of the
emission-line galaxies in this radial bin are within the infall
region.  Note, however, that Figure \ref{interlopers} suggests that
$\sim$30\% of these galaxies have $r_{3D}\gtrsim2.8 r_{200}$.

\subsection{Distribution of Emission Line Galaxies on the Sky}

\begin{figure*}[tb]
\centerline{\epsfxsize=6in\epsffile{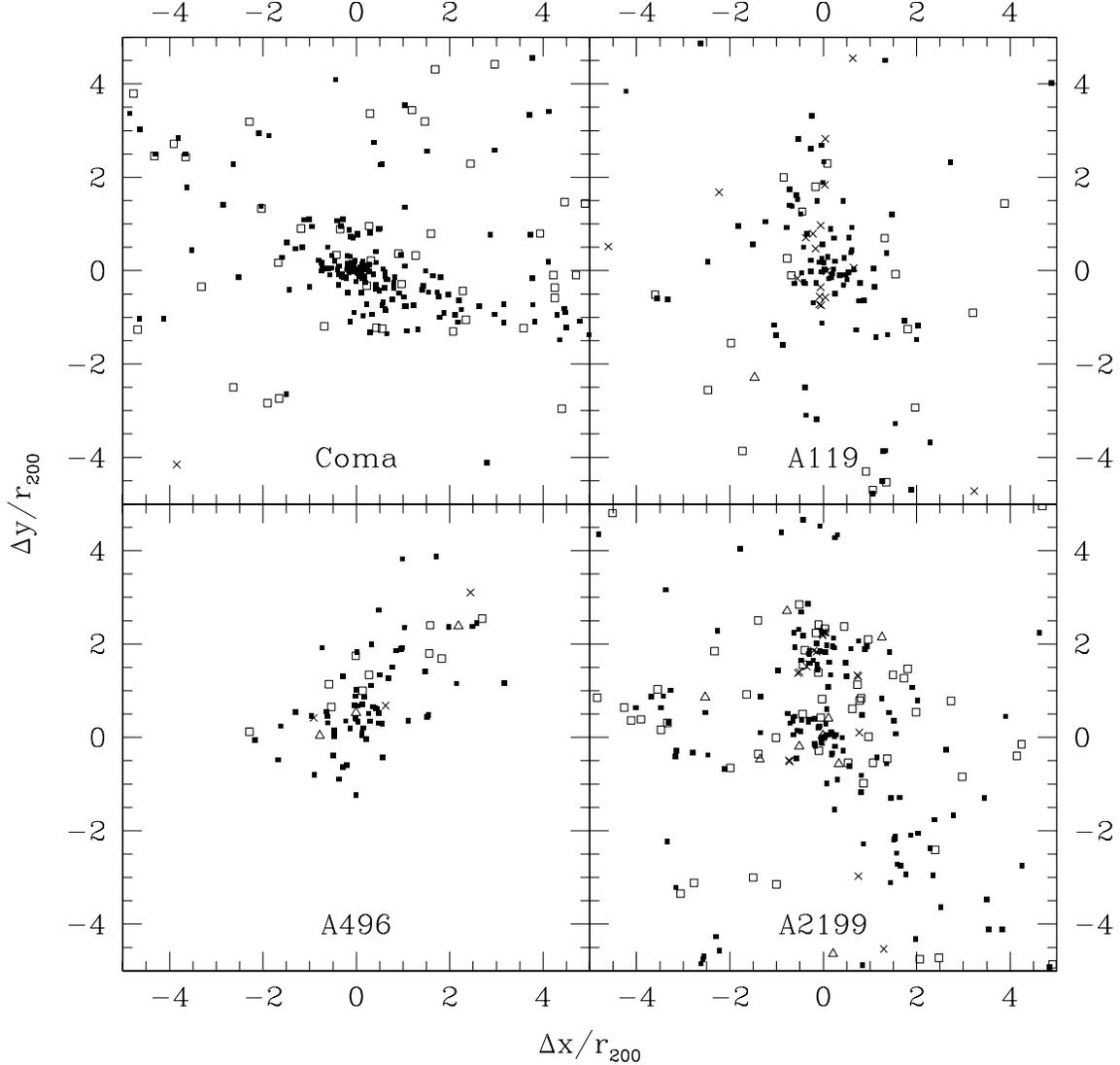}}
\caption{\label{allsky1} Distribution on the sky of different galaxy 
types in four of the CAIRNS clusters.  Solid squares, open squares,
and open triangles show absorption-dominated, star-forming, and AGN
galaxies respectively, while crosses indicate galaxies without FAST
spectra. }
\end{figure*}

\begin{figure*}[tb]
\centerline{\epsfxsize=6in\epsffile{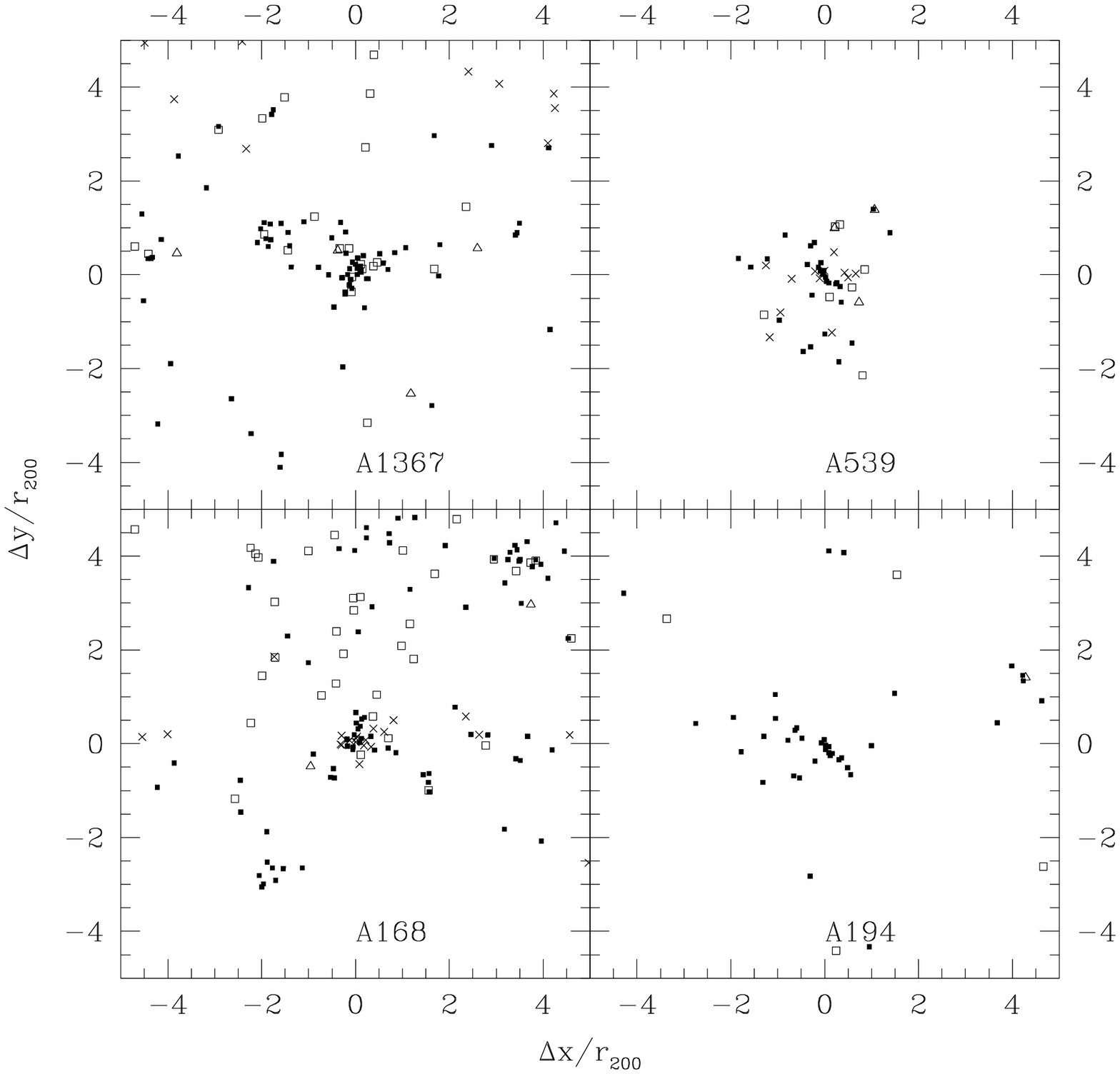}}
\caption{\label{allsky2} Same as Figure \ref{allsky1} for the other 
CAIRNS clusters in the sample. }
\end{figure*}

If emission line galaxies are interlopers from the field, then their
distribution on the sky should not reveal the presence of a cluster.
From the radial distribution of emission line galaxies, we know that
they do not show the cluster as well as the absorption line
galaxies.  Figures \ref{allsky1} and \ref{allsky2} show the
distribution of bright ($M_{K_s}\leq -22.7 + 5 \mbox{log} h$) emission
line and absorption line galaxies on the sky.  Indeed, the emission
line galaxies are much less clustered than absorption line galaxies.
It is evident, however, that the emission-line galaxies trace at least
some of the same structures, although there are large
cluster-to-cluster variations (contrast Coma/A2199/A1367 with
A119/A194).  It is again striking that the kinematics of the
populations are so similar while their spatial distributions differ
dramatically.  Such a situation is possible if the emission-line
galaxies are not in dynamical equilibrium, which would not be
surprising for a population on first or second infall.

\subsection{Spectroscopic Types as a Function of Local Density \label{fracdensity}}

When determining the impact of environment on galaxy properties, it is
critical to distinguish between local and global density.  We now ask
whether the radial dependence of the fraction of galaxies with
emission lines is evident when studying this fraction as a function of
local galaxy density.  Here, we consider the density estimate
$\Sigma_n = n/(\pi D_n^2)$ where $D_n$ is the projected distance to
the $n$th nearest neighbor ($M_K\leq -22.7$).  We use redshifts to
determine cluster membership but not the estimated line-of-sight
separation; for cluster members, $D_n$ is the distance to the $n$th
nearest member ($M_K\leq -22.7$), but for nonmembers, $D_n$ is the
distance to the $n$th nearest galaxy with $\Delta v \leq 1000~\kms$
($M_K\leq -22.7$) regardless of cluster membership.  The distances
$D_n$ may be underestimated for field galaxies close to the infall
regions (where galaxies in the infall region have large peculiar
velocities and are therefore counted as neighbors), but such galaxies
comprise a small fraction of the total field sample.  We restrict the
field galaxies to the redshift range 2000-12000 $\kms$.  We exclude
galaxies in the foreground or background of A119, A168, and A194 from
the field sample because these systems comprise a large fraction of
the foreground or background of the others.  Although not a complete
sample, the absolute magnitude distribution of the field sample is
very similar to the distribution of the cluster/infall sample.  We
exclude all galaxies for which $D_n$ is larger than the distance to
the edge of the survey region.  Because CAIRNS probes rich clusters,
we probe densities up to an order of magnitude larger than studies
based on 2dF and SDSS (which at present contain relatively few rich,
nearby clusters).

\begin{figure*}[tb]
\centerline{\epsfxsize=6in\epsffile{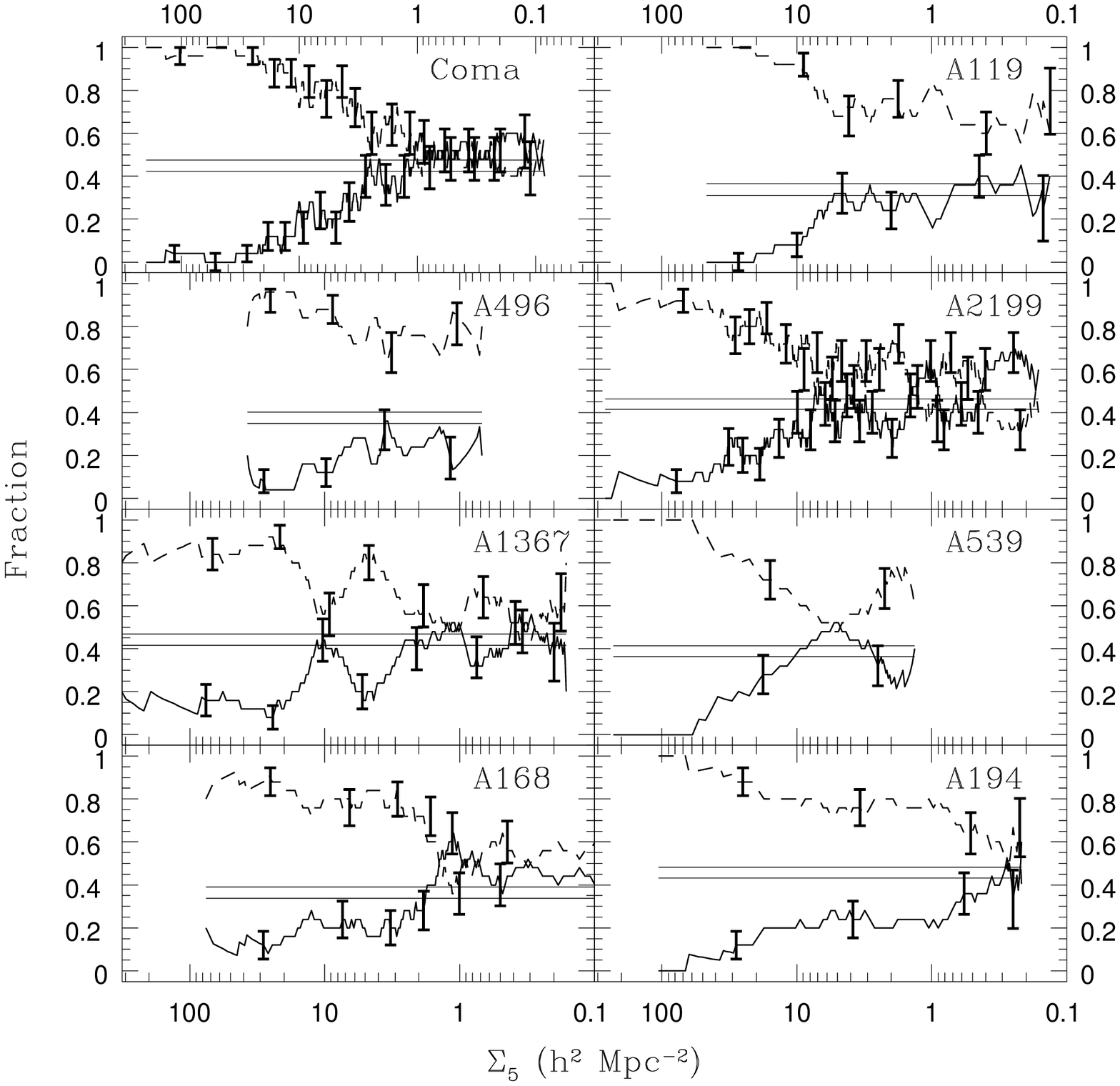}}
\caption{\label{allemfracd5} Fraction of 
galaxies with (solid lines) and without (dashed lines) \ha emission as
a function of local density $\Sigma_5$, the projected density of
galaxies (see text).  The samples for each cluster are selected by
absolute magnitude ($M_{K_s}\leq M_{K_s,comp}$) with a slightly
different limit for each cluster.  As in Figure
\ref{allemfrac}, the lines show moving averages for bins of 25
galaxies with errorbars indicating independent bins.}
\end{figure*}

Figure \ref{allemfracd5} shows the fraction of emission-line galaxies
as a function of local density ($\Sigma_5$).  The trends are very
similar to those in Figure \ref{allemfrac}, demonstrating that local
density is indeed an important characterization of environment.
Figures \ref{allemfrac} and \ref{allemfracd5} are the only ones which
include galaxies fainter than $M_{K_s}=-22.7 + 5 \mbox{log} h$, but
note that the local density estimate uses only galaxies brighter than
this limit (the local densities are therefore directly comparable).
In particular, it is curious that A168 shows a stronger trend with
local density than with distance from the cluster center.  The galaxy
samples used to calculate the emission line fraction differ for the
different clusters, but the local density $\Sigma_5$ is always
calculated only from the bright galaxies ($M_K\leq -22.7$).

\begin{figure*}[tb]
\centerline{\epsfxsize=5in\epsffile{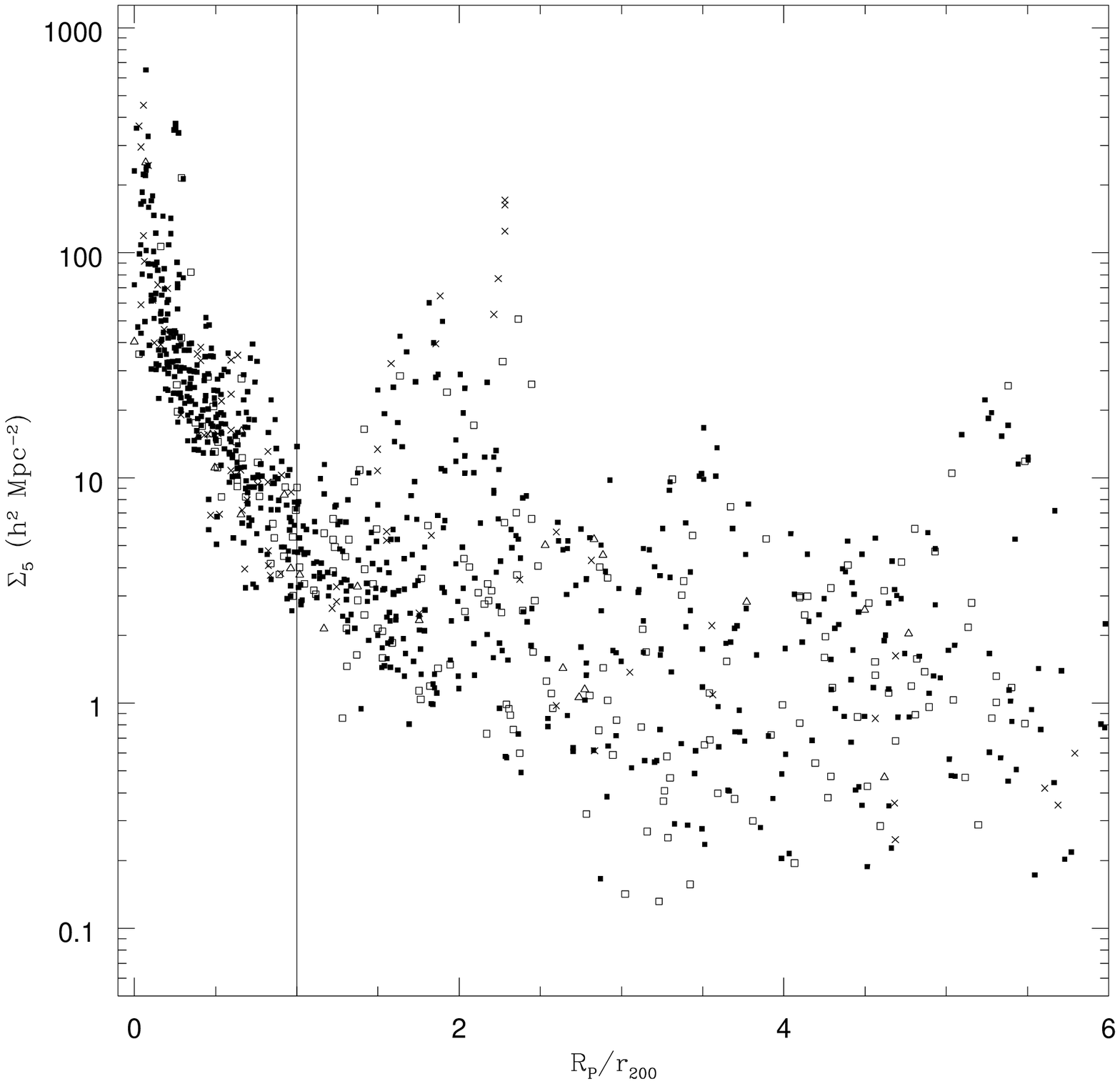}}
\caption{\label{s5r200} Local density $\Sigma_5$ versus clustercentric
 radius of bright ($M_{K_s}\leq -22.7$) galaxies in all CAIRNS
clusters.  Solid squares, open squares, and open triangles show
absorption-dominated, star-forming, and AGN galaxies respectively,
while crosses indicate galaxies without FAST spectra.  The vertical
line at $R_{200}$ separates the virial region from the infall region.}
\end{figure*}

We now test whether local or global environment is more important in
determining the star formation properties of the inner parts of
galaxies.  Figure \ref{s5r200} shows the local density $\Sigma_5$ as a
function of clustrocentric distance.  These two parameters are
significantly correlated inside $r_{200}$, but beyond $r_{200}$ (the
infall regions), there is a wide range of $\Sigma_5$ at a given value
of $R_p$.  Thus, if clustrocentric radius is more important than local
density, the emission fraction of galaxies in the infall region should
exhibit little or no dependence on $\Sigma_5$.

\begin{figure*}[tb]
\centerline{\epsfxsize=5in\epsffile{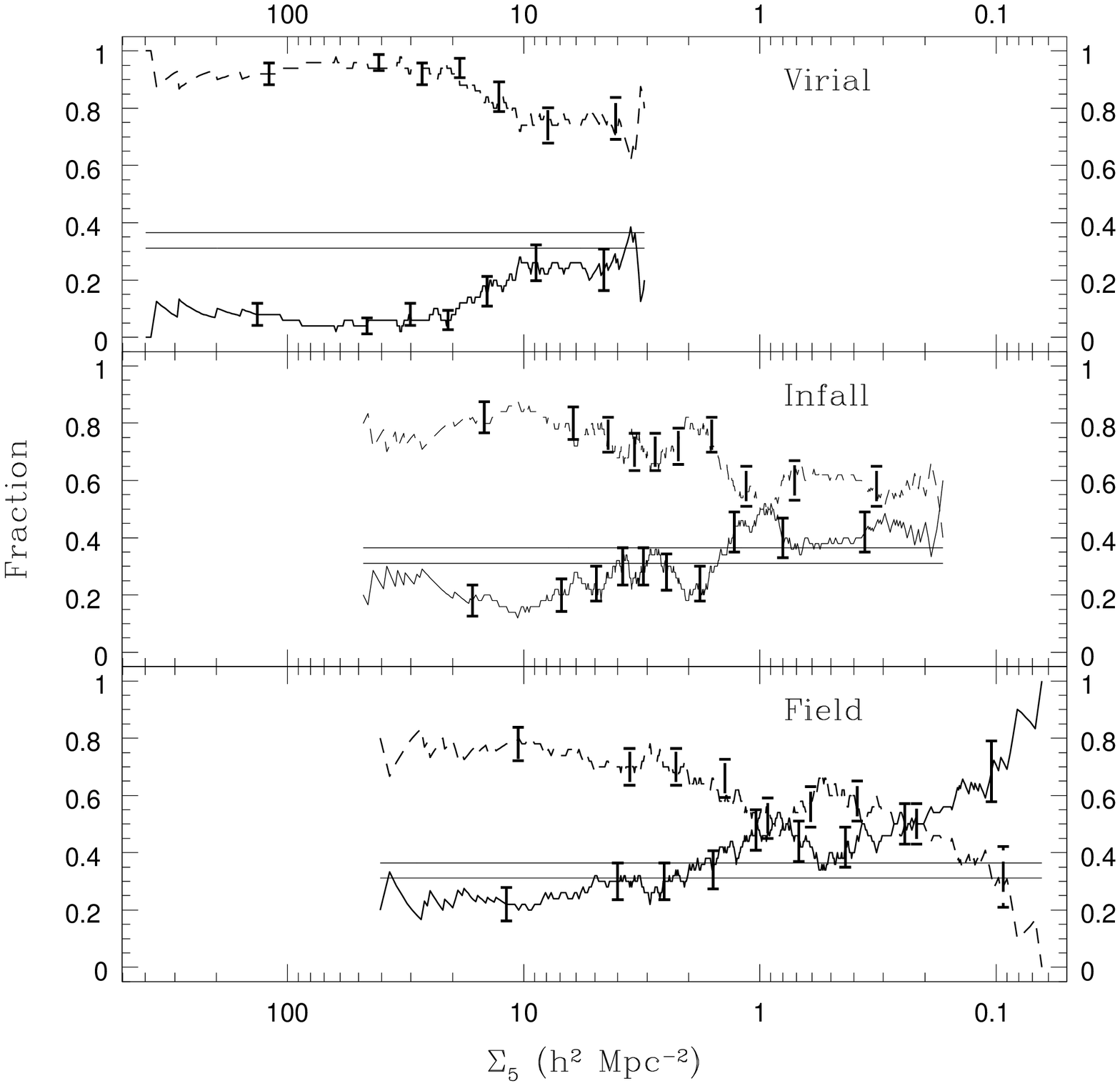}}
\caption{\label{allemfracd5vif} Fraction of bright 
($M_{K_s}\leq -22.7$) galaxies with (solid lines) and without (dashed
lines) \ha emission as a function of local density $\Sigma_5$, the
projected density of galaxies (see text).  As in Figure
\ref{allemfrac}, the lines show moving averages for bins of 25
galaxies with errorbars indicating independent bins.}
\end{figure*}

We separate the galaxies into three types of global environment in
Figure \ref{allemfracd5vif}.  The three panels show the emission line
fractions of bright galaxies inside the virial regions ($R_p\leq
R_{200}$), inside the infall regions ($1<R_p/R_{200}\leq 5$) and
``field'' galaxies (all nonmembers with the absolute magnitude and
redshift cutoffs described above).  At fixed local density, the
emission line fraction is independent of global environment (virial
region, infall region, or field).  This result demonstrates that local
density is the primary environmental factor in determining the current
star formation properties of galaxies.  This conclusion is consistent
with previous investigations of galaxies in and near clusters
\citep{balogh97,diaferio01,ellingson01,lewis02,gomez03,treu03,balogh03,gray04,tanaka04},
and the emission fractions agree well with \citet{balogh03}.  We
extend this relation to densities larger by a factor of a few than
those probed in \citet{balogh03}.  The emission fraction of galaxies
in the virial region declines significantly at densities
$\Sigma_5\gtrsim 20 h^2 \mbox{Mpc}^{-2}$, a regime previously unprobed
(Figure \ref{allemfracd5vif}).

The strong trend in the field sample (bottom panel of Figure
\ref{allemfracd5vif}) is perhaps surprising, as most previous studies
restrict their samples to galaxies in and near clusters and groups
where the local density is thought to be more reliable.  This
dependence probably indicates that there are some galaxy groups in the
field sample (we only exclude galaxies in cluster infall regions from
the field sample, so these galaxies should sample a variety of
environments).  \citet{bcarter} find a similar trend in the 15R survey
(where galaxies are not selected by environment), again indicating the
importance of local density.  This dependence complicates the
determination of the emission line fraction of field galaxies.  A
similar result was noted by \citet{mateus04}, who found that the
environmental dependence of the emission-line fraction continues to
very low densities.

\section{Environmental Dependence of Star Formation Rates \label{distribha}}

In the previous section we showed that the fraction of galaxies with
current star formation is dramatically reduced in and around clusters.
To find the mechanism driving this reduction in star formation rates
we now analyze the star formation rates of cluster galaxies with
current star formation.  For instance, if the reduction in star
formation rates is a gradual process such as starvation, we expect the
distribution of star formation rates in cluster galaxies to be skewed
to lower rates than in field galaxies (of similar luminosities) with
current star formation.

\begin{figure*}[tb]
\centerline{\epsfxsize=6in\epsffile{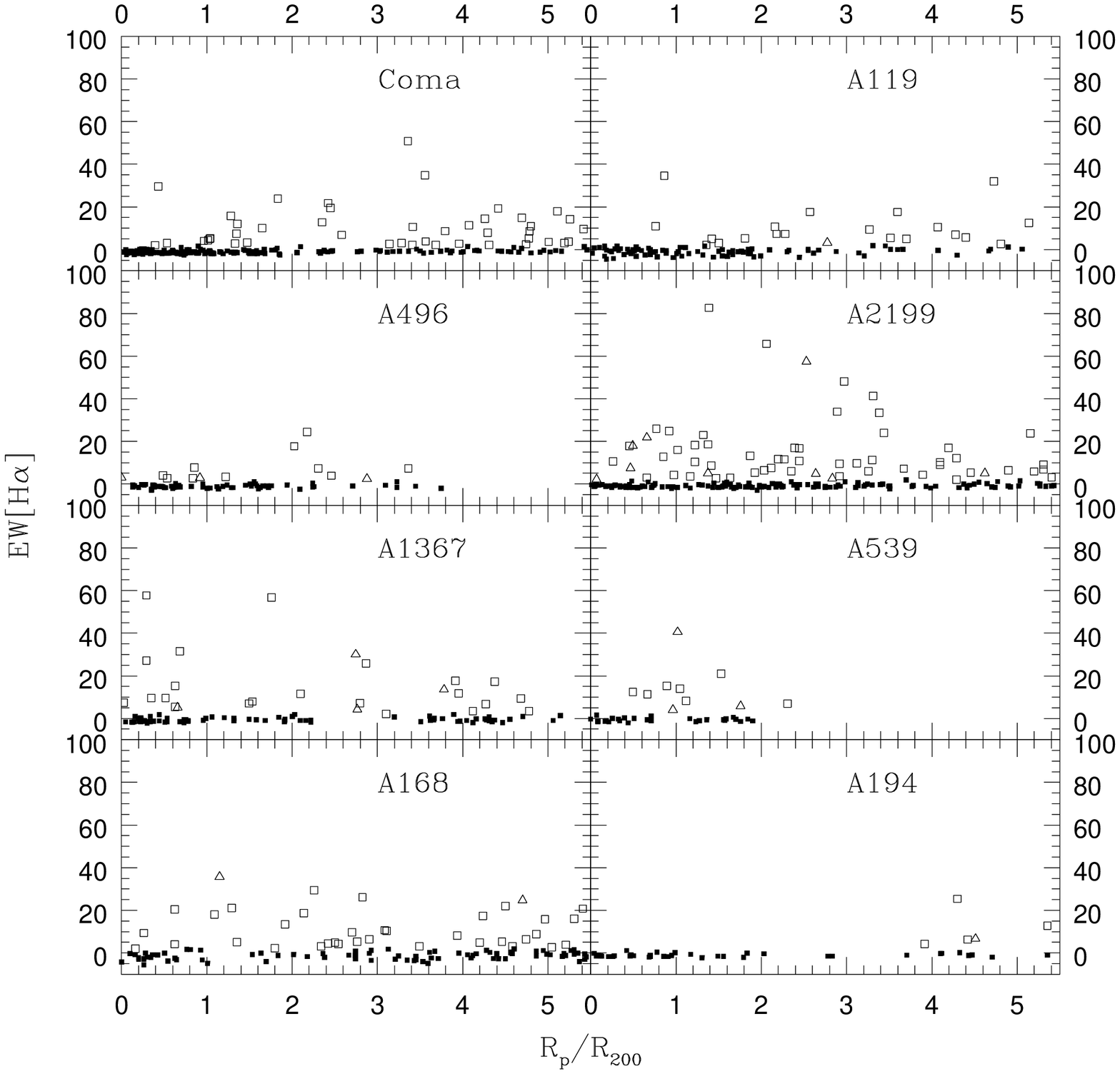}}
\caption{\label{alleqw}  \ewha ~versus clustrocentric radius.  
 Solid squares, open squares,
and open triangles show absorption-dominated, star-forming, and AGN
galaxies respectively.}
\end{figure*}

Figure \ref{alleqw} shows \ewha ~versus clustrocentric radius for
galaxies in the $K_s$ selected samples.  Different points show
galaxies with absorption lines, AGN, and star formation.  Naively, one
might expect that the distribution would be contained in a triangular
envelope with small EWs at small radii.  Instead, galaxies with strong
emission lines are present at all radii.  These galaxies may be recent
arrivals which have not yet been stripped or they may be infall
interlopers ($\S \ref{radial}$).

\begin{figure*}[tb]
\centerline{\epsfxsize=6in\epsffile{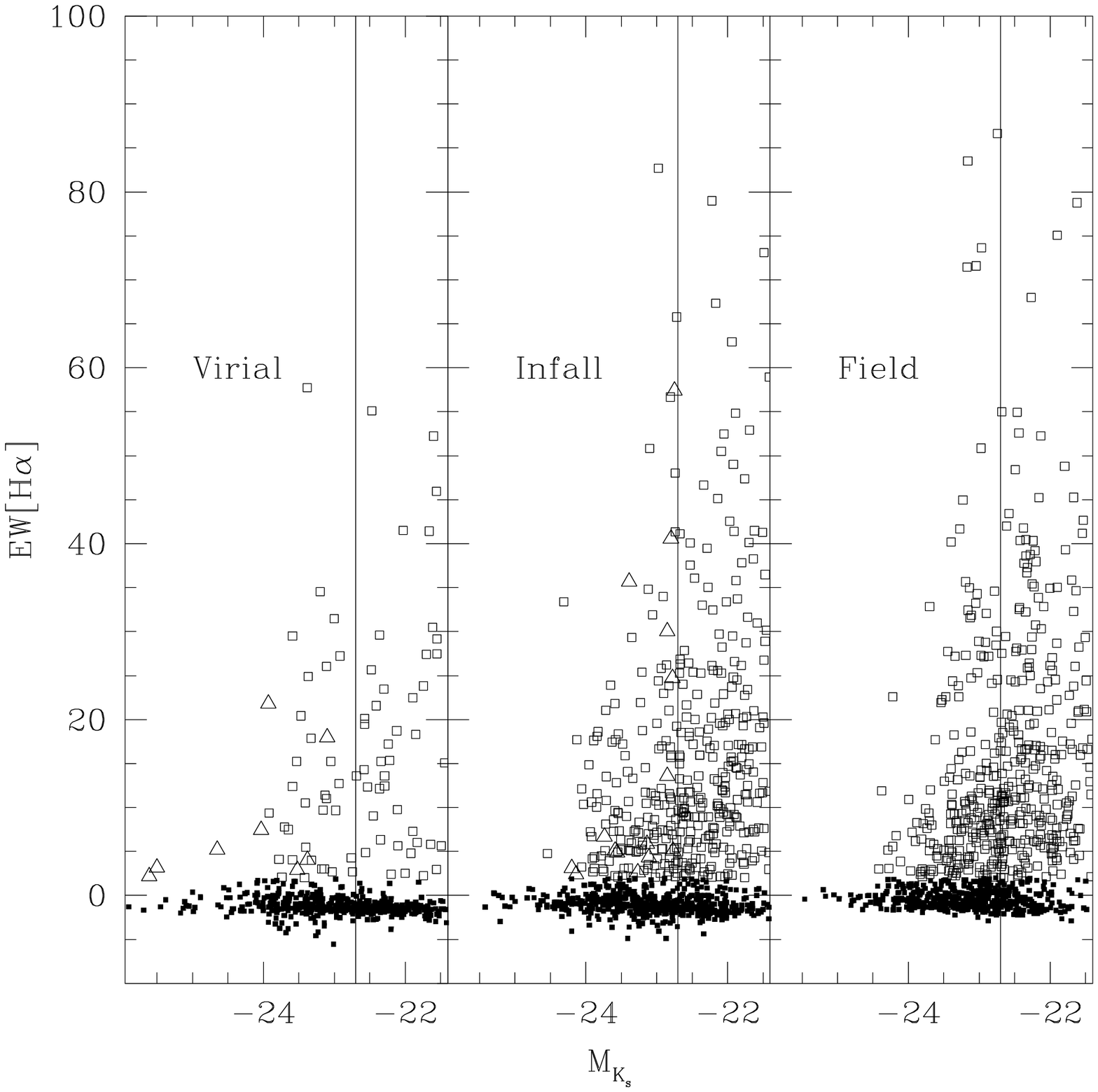}}
\caption{\label{fieldcomp}  \ewha ~versus absolute magnitude (from 
left to right) for galaxies in cluster virial regions, infall regions,
and the field.   Solid squares, open squares,
and open triangles show absorption-dominated, star-forming, and AGN
galaxies respectively.  Vertical lines indicate the spectroscopic completeness
limit of the CAIRNS clusters.}
\end{figure*}

\begin{figure}[tb]
\plotone{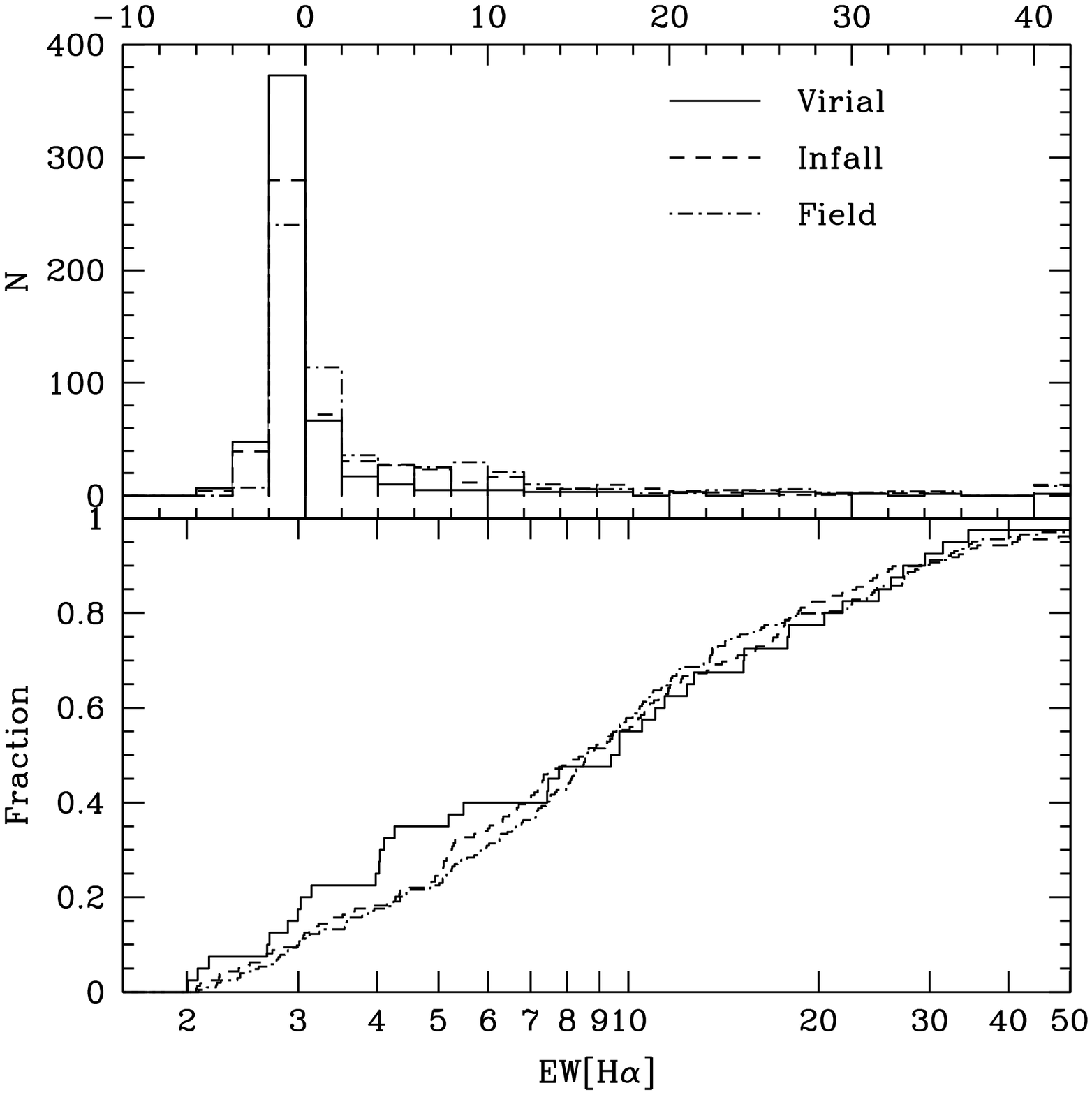} 
\caption{\label{fieldcomp2}  (Top panel) Distribution of \ewha~in different 
environments.  Solid, dashed, and dash-dotted lines indicate galaxies in virial regions, infall regions, and the field respectively.  Histograms have been normalized to have equal areas.  (Bottom panel) Cumulative distributions of \ewha~for galaxies with \ewha$\geq$2\AA.}
\end{figure}

Figure \ref{fieldcomp} shows the distribution of \ewha ~versus
absolute magnitude for galaxies in three types of environment: cluster
virial regions, cluster infall regions, and the field.  Consistent
with many other studies, fainter galaxies tend to have stronger
emission lines.  CAIRNS virial regions contain few galaxies with
strong emission lines.  The distribution of \ewha ~for galaxies in the
CAIRNS infall regions, however, is very similar to that in the field.
It is curious that the only emission-line galaxies brighter than
$M_{K_s}=-25$ in {\it any} environment are the cD galaxies of A496 and
A2199, both of which are AGN.  Figure \ref{fieldcomp2} shows
histograms of \ewha ~in these three environments both for all bright
galaxies and for only bright galaxies with \ewha$>$2\AA.  We use K-S
tests to test whether the distributions of \ewha ~in the cluster,
infall, and field samples are different.  The overall bright galaxy
samples in the three environments are definitely different (a K-S test
distinguishes them with more than 99.9\% confidence).  Among bright
galaxies with emission lines, however, we find no significant
differences (at the 90\% confidence level) between samples with the
same absolute magnitude limits.  Similarly, we find no differences
among the three environments for galaxies with moderately strong
(\ewha$>$10\AA) emission lines.  This weak dependence of \ewha ~on
environment was first observed by \citet{bcarter} for galaxies in all
environments from the 15R survey, and this weak dependence has been
confirmed by \citet{balogh03} and \citet{tanaka04}, who find similar
results in a study of galaxies in both 2dFGRS and SDSS.

One potential explanation of the similarity of the distributions of
\ewha ~in different environments is that changes in star formation
rates occur abruptly, perhaps due to ram pressure stripping or strong
tidal interactions \citep{vandokkum98,balogh03}.  Galaxies with star
formation that is truncated abruptly should show evidence of a
population of A stars, i.e., K+A or E+A galaxies
\citep[e.g.,][]{kauffmann04}.  We search for galaxies with strong
Balmer absorption (\ewhd$>$5\AA) and find two candidates among
$\approx$1000 bright galaxies and one definite E+A in A168 at 0.3 mag
below our minimum luminosity.  Curiously, this galaxy lies just
23$\kpc$ in projection from and at the same redshift as a bright
galaxy in our sample that is 2.15 mag brighter than the E+A galaxy.
Both of the bright candidates are consistent with \ewhd$\leq$5\AA~at
the 1-$\sigma$ level and one of them has a SDSS spectrum which shows
emission lines.  The frequency of E+As among bright galaxies in nearby
clusters is therefore $\lesssim$0.1\%, crudely consistent with other
studies of local galaxies in all environments
\citep{1996ApJ...466..104Z,2004ApJ...602..190Q}.  The FAST spectra have
lower S/N at blue wavelengths.  It is thus possible that we miss some
of these galaxies due to low S/N.  A definitive search would require
followup spectra with higher S/N.  However, other investigators have
found that Coma contains no bright post-starburst/post-star-forming
galaxies, unlike clusters at moderate redshift
\citep{1997AJ....113..492C,poggianti04}, although Coma does contain
some faint post-starburst galaxies \citep{1996AJ....111...78C}. 

\citet{kauffmann04} show that the relations between star formation
tracers do not depend on environment, despite the fact that the
tracers are sensitive to different timescales of star formation.  They
conclude that the only way to maintain these relations is for
cessation of star formation to occur on very long timescales
($\gtrsim$1 Gyr).  However, this conclusion is based on a specific
star formation history where star formation is constant for 10 Gyr
($\mbox{log}(SFR/M_*)=-10$ where $M_*$ is the total stellar mass of
the galaxy) after which star formation halts abruptly.  This model is
reasonable for all emission-line galaxies except those with
$\mbox{log} ~M_*\gtrsim 10$ \citep{brinchmann04}.  However, our sample
and that of \citet{balogh03} consists mostly of galaxies with these
high masses.  \citet{brinchmann04} find that galaxies with $\mbox{log}
~M_*> 10.5$ typically have $\mbox{log}(SFR/M_*)\leq-10.5$, indicating
that their SFRs were greater in the past.  Detailed modeling of the
star formation histories of more massive galaxies may resolve this
discrepancy.

The models shown by \citet{kauffmann04} reach the SF relations of
normal galaxies after $\sim$1.5 Gyr.  Thus, if cessation of star
formation (by any mechanism) occurs largely at moderate or high
redshifts \citep[an alternative suggested by][]{balogh03}, these
galaxies would have sufficient time to reach the SF relations observed
for local galaxies.  This type of solution points towards the
significance of the initial conditions of galaxy formation and thus
blurs the line between effects related to the initial conditions of
galaxy formation and effects related to subsequent environmental
evolution.  Detailed studies of \ha properties at moderate redshift
should help resolve this conflict.  For instance, \citet[][]{kodama04}
use narrowband imaging and photometric redshifts to analyze one
cluster at $z\approx$0.4 and find results similar to those we find for
clusters in the nearby universe: the fraction of galaxies with \ha
emission depends strongly on local density, but the \ha luminosity
function does not depend strongly on environment.  These results,
particularly if they are true generally for clusters at moderate
redshift, strengthen the case for rapid truncation of star formation.
Curiously, the star formation-density relation in this cluster is much
stronger than the morphology-density relation discussed in
\citet{treu03}, suggesting that star formation is more sensitive than
morphology to environment.

Alternatively, there may be competing processes affecting star
formation, e.g., reduced star formation due to stripping of the halo
gas reservoir but enhanced circumnuclear star formation due to tidally
induced starbursts.  Because fiber spectra and our FAST spectra are
dominated by light from the inner few kpcs of galaxies, these spectra
may have larger measured \ewha ~than integrated spectra (see the
discussion in $\S \ref{discuss}$).  An objective prism study of
cluster galaxies suggests that cluster galaxies have a higher
incidence of tidally induced circumnuclear starbursts than field
galaxies \citep{moss00}.  Detailed modeling would be required to see
whether such competing processes would cause galaxies to deviate
significantly from the relations among SF tracers.  Finally, the
importance of projection effects especially for emission-line galaxies
($\S \ref{radial}$) suggests that many, perhaps most emission-line
galaxies lie at much lower spatial densities.  The similarity of the
two SFR distributions would then be a consequence of drawing from the
same parent distribution of local spatial density and artificially
large projected densities.

\begin{figure*}[tb]
\centerline{\epsfxsize=6in\epsffile{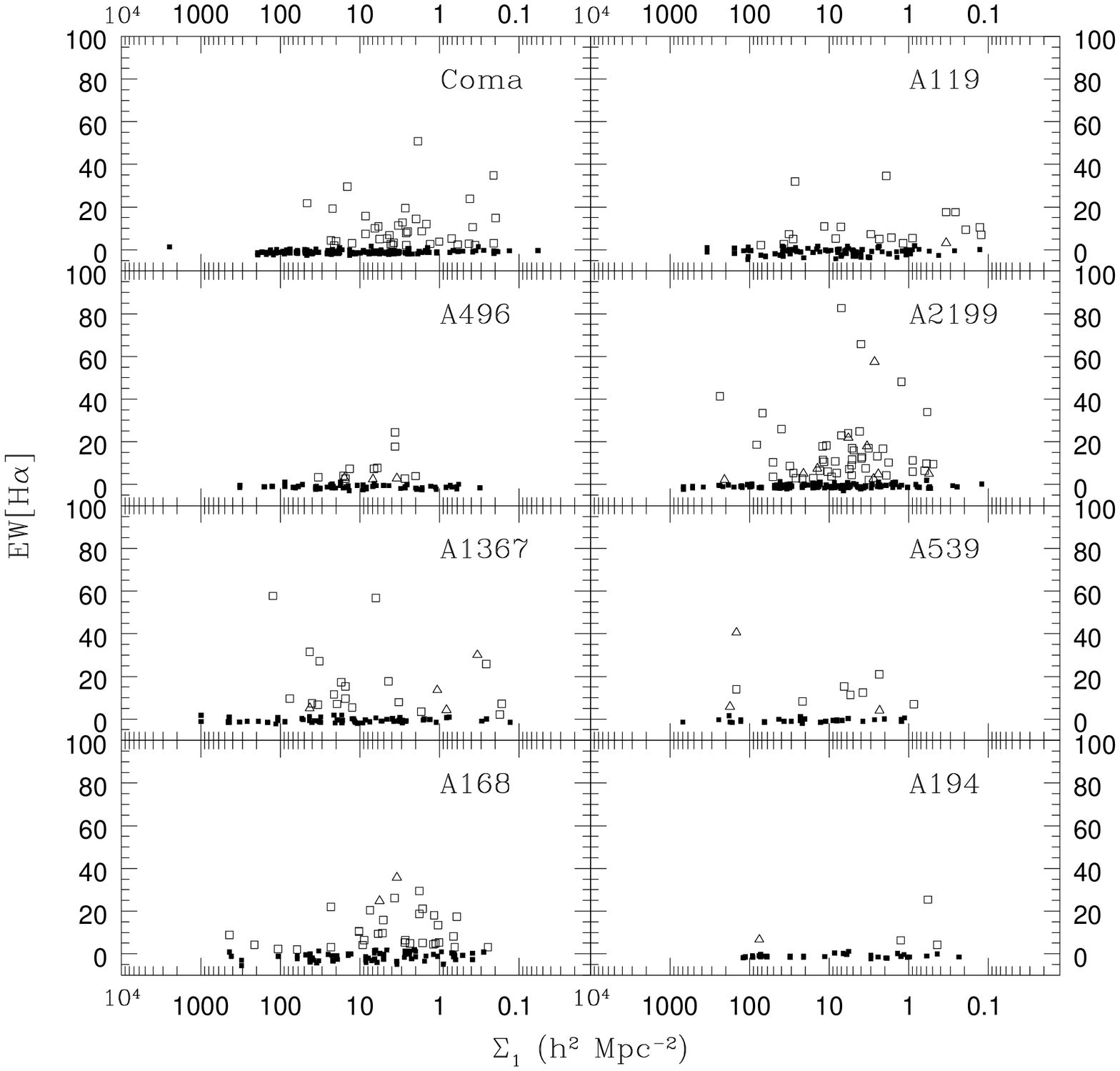}}
\caption{\label{alleqwd1}  \ewha ~versus local density $\Sigma_1$ (computed from the nearest neighbor distance) for the CAIRNS clusters.   Solid squares, open squares,
and open triangles show absorption-dominated, star-forming, and AGN
galaxies respectively.}
\end{figure*}

\begin{figure*}[tb]
\centerline{\epsfxsize=6in\epsffile{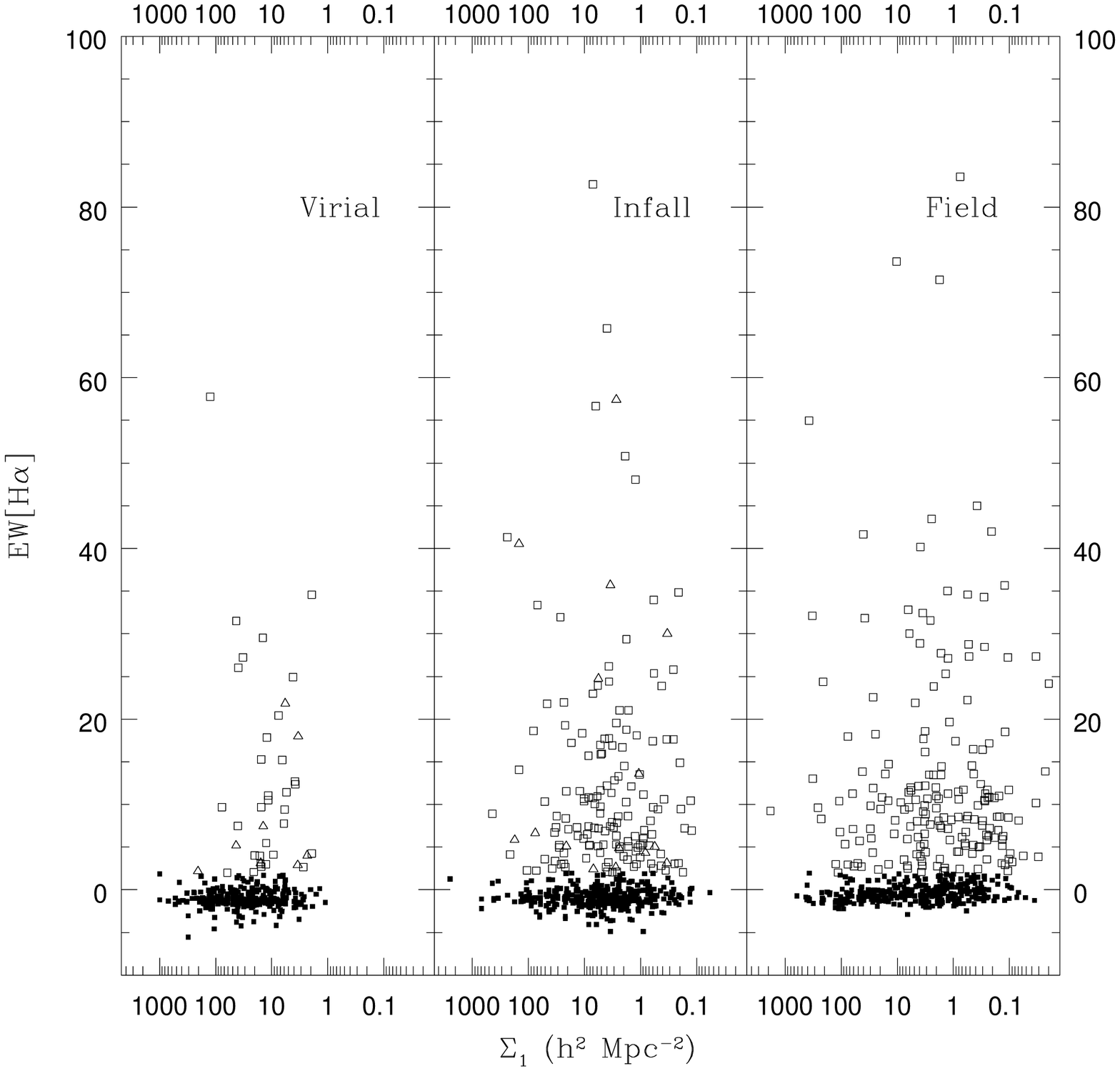}}
\caption{\label{alleqwd1b}  \ewha ~versus nearest neighbor distance $\Sigma_1$ (from left to right) for galaxies in cluster virial regions, infall regions, and the field.   Solid squares, open squares,
and open triangles show absorption-dominated, star-forming, and AGN
galaxies respectively.}
\end{figure*}

\begin{figure*}
\centerline{\epsfxsize=6in\epsffile{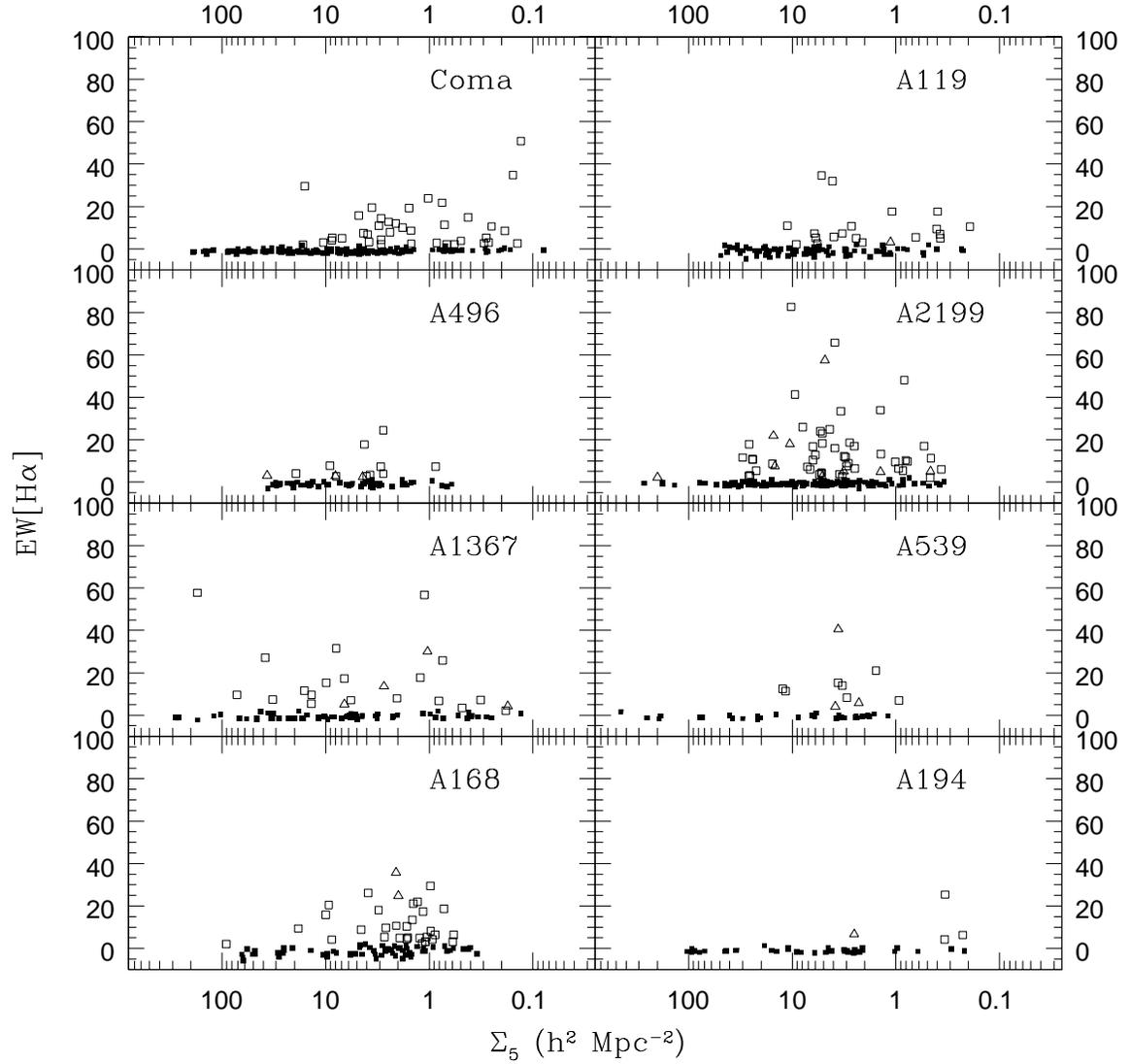}}
\caption{\label{alleqwd5}  \ewha ~versus $\Sigma_5$ for the CAIRNS clusters.   Solid squares, open squares,
and open triangles show absorption-dominated, star-forming, and AGN
galaxies respectively.}
\end{figure*}

\begin{figure*}[tb]
\centerline{\epsfxsize=6in\epsffile{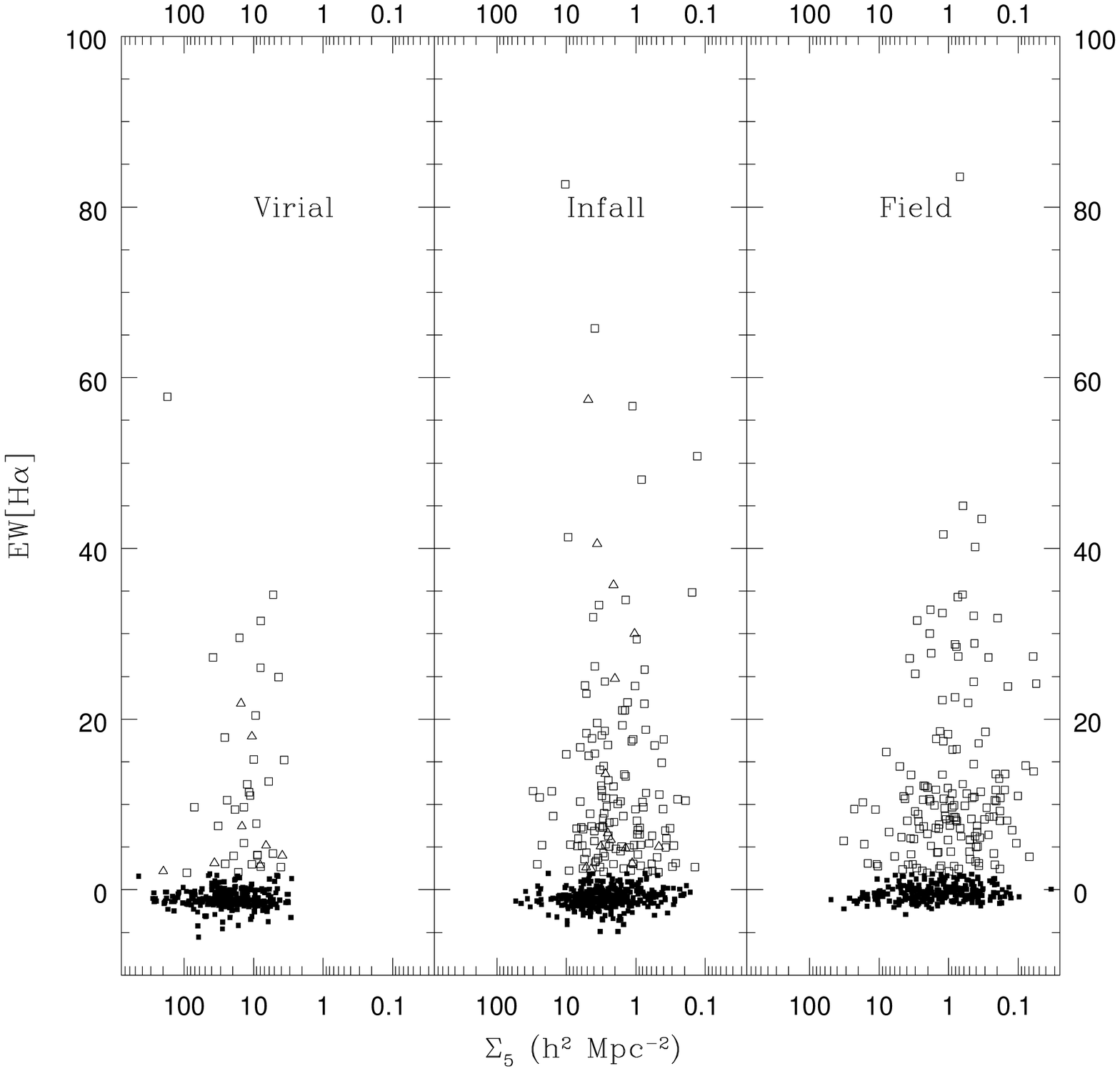}}
\caption{\label{alleqwd5b}  \ewha ~versus $\Sigma_5$ (from left to right) for galaxies in cluster virial regions, infall regions, and the field.   Solid squares, open squares,
and open triangles show absorption-dominated, star-forming, and AGN
galaxies respectively.}
\end{figure*}

We now test other density estimators to determine whether suppression
of star formation might be correlated with a more local density
estimate.  We consider four density estimates: the distance $D_n$ to
the $n$th nearest neighbor ($M_K\leq -22.7$) for $n=1$, 5, and 10
(defined in $\S \ref{fracdensity}$), and the number of galaxies
$N_{1.1}$ within a circle of radius 1.1$\Mpc$.  Figure \ref{alleqwd1}
shows \ewha ~as a function of the distance to the nearest (bright)
neighbor in the CAIRNS clusters.  We exclude all galaxies with $D_1$
is larger than the distance to the edge of the survey region.  Figure
\ref{alleqwd1b} shows this relation in three environments: virial
regions, infall regions, and the field.  Because of the large
(projected) densities in the virial regions, there are few
``isolated'' galaxies in the virial regions with large values of
$D_1$.  Also, ``close pairs'' (i.e.,$D_1\lesssim 50~\kpc$) in virial
regions contain few galaxies with large \ewha ~compared to close pairs
in infall regions and the field.  The close pairs in the infall and
field environments with large \ewha ~($>$40\AA)are likely tidally
induced starbursts \citep{2000ApJ...530..660B}.  Apart from this
difference, the distribution of \ewha ~versus $D_1$ appears to be very
similar in the three environments.  Figures \ref{alleqwd5} and
\ref{alleqwd5b} show the similar distributions of \ewha ~versus $D_5$.
We find similar distributions of \ewha ~versus $D_{10}$ and $N_{1.1}$
which we omit for brevity.  There are no obvious differences in these
distributions among the three types of environment.  We confirm this
quantitatively by comparing the distributions of \ewha ~for cluster
and infall region galaxies above and below the median value of $D_1$
and $D_5$ with a K-S test; the tests show no differences at the 95\%
confidence level.  We similarly find no differences in the
distributions of \ewha ~for galaxies in the most and least dense
quartiles.
 
\begin{figure}[tb]
\plotone{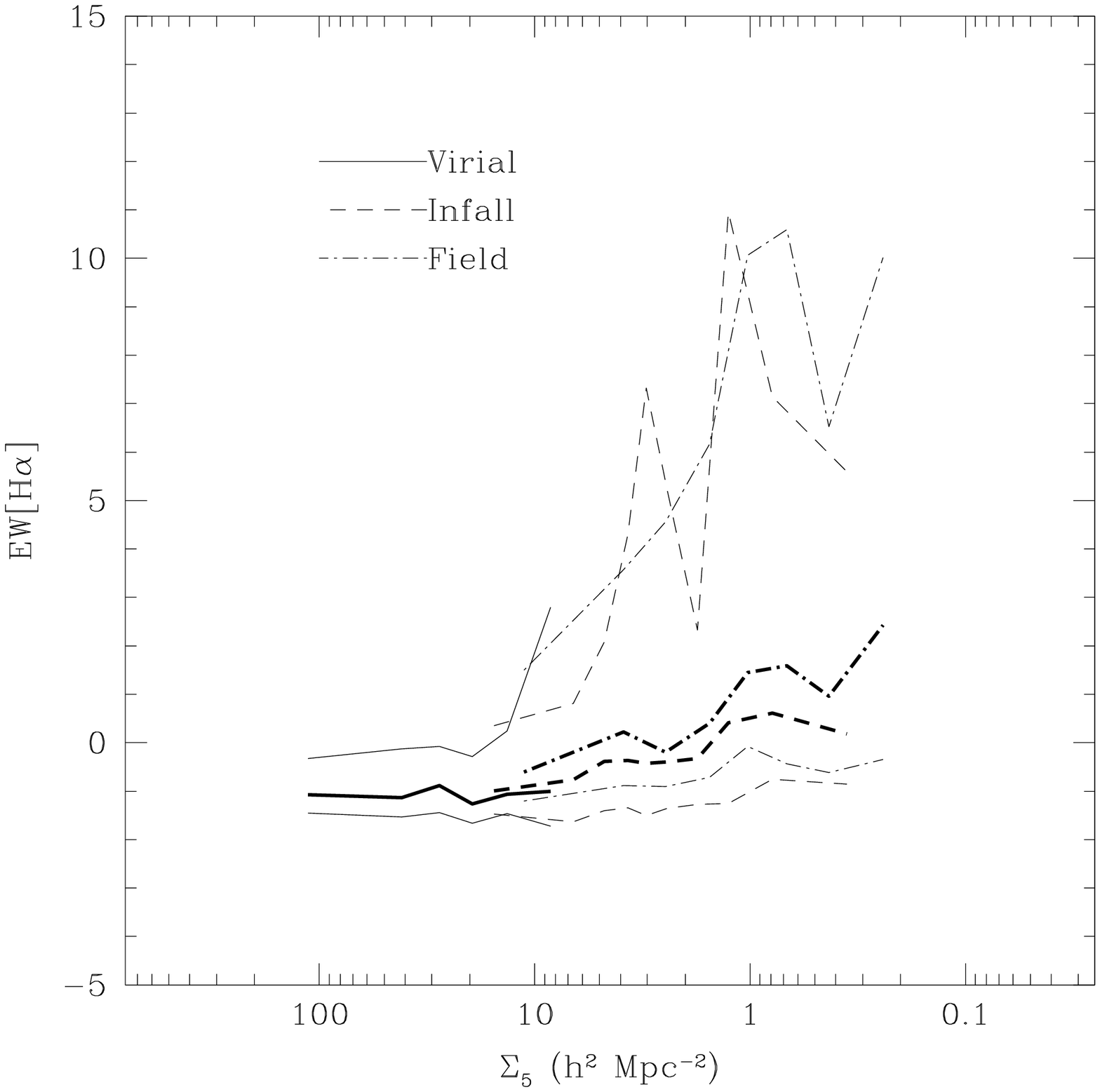} 
\caption{\label{quarall}  Medians (thick lines) and quartiles (thin lines) of the distributions of \ewha ~versus local density $\Sigma _5$ in the three global environment types. }
\end{figure}

Several authors have discussed the median and quartiles of the
ditribution of \ewha~ as a function of local density
\citep{gomez03,balogh03,tanaka04}.  They find evidence for a ``break''
in the distributions at a local density of $\Sigma_5\approx 2 h^2
Mpc^{-2}$.  \citet{balogh03} notes that this break reflects the local
density where the emission-line fraction drops below 25\% or 50\%.
Figure \ref{quarall} shows the quartiles of the \ewha~ distribution as
a function of local density for the CAIRNS galaxies in the three
global environment types (virial regions, infall regions, and field).
We confirm that the upper quartile shows a break at $\Sigma_5\approx
1-2 h^2 Mpc^{-2}$.  The quartiles of \ewha~ for the $K$-selected
CAIRNS galaxies are similar to those of the red-selected SDSS galaxies
in \citet{balogh03} and \citet{tanaka04} and smaller than those of the
blue-selected 2dFGRS galaxies in \citet{balogh03}.  A striking feature
of Figure \ref{quarall} is that the quartiles show a similar
dependence on $\Sigma_5$ in all three environment types.  This result
further strengthens the case for the primacy of local density in
determining galaxy properties.

\section{Discussion \label{discuss}}

We confirm and extend recent results on the environmental dependence
of star formation.  Our study utilizes a complete, near-infrared
sample of cluster galaxies in clusters with robustly determined
parameters.  Our near-infrared selection yields a sample that
is much closer to a sample selected by stellar mass than samples
selected at optical wavelengths.  In optically selected samples, the
stellar masses of star forming galaxies near the survey limit are
smaller than the stellar masses of galaxies near the survey limit
without star formation.  Thus, near infrared selection significantly
reduces this potential systematic bias.

Our galaxy spectra are obtained with long-slit observations, which are
relatively insensitive to aperture bias \citep{bcarter}.  However, our
spectra are still heavily weighted towards the nuclear regions of
galaxies relative to true integrated spectra
\citep[e.g.,][]{2000ApJS..126..331J}.  This weighting is important to
remember when comparing our results to, e.g., detailed studies of
Virgo galaxies which show ongoing ram pressure stripping and evidence
of truncation of star-forming disks
\citep{koopmann04b,2004AJ....127.3361K}.  

Compared to fiber surveys like 2dFGRS and SDSS, the CAIRNS sample
avoids incompleteness due to fiber collisions.  This completeness is
very important when studying the properties of cluster galaxies with
current star formation because these galaxies are quite rare.  

Our long-slit spectra are dominated by light from the inner few kpcs
of galaxies, but they sample some light from the outer parts of
galaxies.  We therefore expect our results to be more robust to
aperture bias than fiber surveys.  The consistency of our results with
previous investigations suggests that aperture bias and spectroscopic
completeness do not cause significant systematic biases in 2dF and
SDSS.

A recent study of Virgo spirals \citep{koopmann04b} suggests an
important caveat to the impact of aperture bias on our results.
\citet{koopmann04b} find that many Virgo spirals have truncated \ha
disks \citep[see also][]{vogt04b}.  Because these truncated disks are
not often found in samples of field galaxies, estimates of aperture
bias based on field samples \citep{bcarter,kewley05} may not apply to
cluster samples.  We attempt to quantify the impact of truncated
spirals on our results.  \citet{koopmann04b} list \ewha both inside
$r_{24}$ (the R-band isophote of 24 mag arcsec$^{-2}$) and inside $0.3
r_{24}$ (their Table 3).  From this sample, we find that the ratio
EW(inner)/EW(total) is a factor of $\approx$2.2 larger for galaxies
classified as truncated versus all other galaxies.  To estimate the
{\em maximum} impact this truncation would have on our results, we
assume that {\em all} galaxies inside $R_{200}$ are truncated and that
{\em no} galaxies outside $R_{200}$ are truncated.  Then, a K-S test
would detect a difference between the emission-line populations
inside/outside $R_{200}$ (Figure \ref{fieldcomp2}) at the 99.8\%
confidence level.  However, some galaxies inside $R_{200}$ are
probably infall interlopers and some galaxies outside $R_{200}$ are
backsplash galaxies.  Thus, this estimate is a conservative upper
limit on the difference in the two (projected) populations.

Our spectroscopic samples are limited to relatively bright galaxies
($M\lesssim M^*+1$).  Recently, \citet{tanaka04} have shown that
slightly fainter galaxies ($1<M-M^*<2$) in SDSS show different
environmental dependencies than bright ($M\leq M^*+1$) galaxies.  The
properties of galaxies in the luminosity bin $1<M-M^*<2$ and those of
dwarf galaxies are not constrained by our observations.

We measure star formation rates from \ewha, similar to many other
investigators
\citep{bcarter,lewis02,gomez03,balogh03,mateus04,tanaka04}.  In
assuming \ewha ~is directly proportional to SFR (normalized by galaxy
luminosity), we assume that the extinction to continuum and to H\II
~regions are identical.  If this assumption is incorrect
\citep{calzetti01}, then Balmer decrements are necessary to obtain
accurate SFR estimates.  However, our conclusion that the distribution
of \ewha ~{\it among galaxies with current star formation} does not
depend on environment ($\S \ref{distribha}$) holds unless the relative
extinctions of continuum and H\II ~regions depend on environment.

\section{Conclusions \label{conclusions}} 

The environmental dependence of galaxy properties can be decomposed
into two effects: the {\em fraction} of galaxies with current star
formation and the {\em distribution of star formation rates among
galaxies with current star formation}.  Infall regions (where galaxies
are infalling onto the main cluster but have not yet reached
equilibrium) provide a unique probe of the size scale on which
environment is important: infall regions are overdense on scales of
5-10$\Mpc$ (because they are near a cluster) but they contain a wide
range of local densities measured on scales of roughly 1$\Mpc$.

The CAIRNS sample confirms that the fraction of galaxies with current
star formation is suppressed both inside and outside the virial
regions of rich clusters.  For the first time, we show this effect in
individual clusters as well as cluster-to-cluster variations in the
effect.  One cluster, A2199, contains X-ray groups at large radii.
The emission fraction profile shows a dip at the radius of these
groups, indicating that they are sufficiently massive to influence
galaxy properties.  More generally, the emission-line fraction depends
strongly on local projected galaxy density, and this dependence is the
same in different types of global environment (virial regions, infall
regions, and the field).  These results together suggest a limited
role for cluster-specific mechanisms.  We extend this spectral
type-density relation to higher local densities than probed by 2dF and
SDSS (which at present contain relatively few rich, nearby clusters).
Our results indicate that previous studies based on 2dF and SDSS data
have not been biased by systematic effects due to fiber spectra,
aperture bias, or incompleteness in dense regions.

The radial distribution of both emission-line and absorption-line
galaxies closely follow NFW profiles, but with very different
concentrations, $c\approx$4.3 and 0.8 for absorption and emission-line
galaxies respectively.  Both concentrations are smaller than the mass
profiles in Paper I.  \citet{lin04} reached similar conclusions about
the relative distributions of galaxies and dark matter.  From these
profiles we can predict the rate of ``infall interlopers,'' galaxies
with $R_p\leq r_{200}$ but $r_{3D}>r_{200}$.  At least 20\% of
absorption line galaxies and 50\% of emission line galaxies are infall
interlopers.  Departures from spherical symmetry would likely increase
the fraction of emission-line galaxies which are infall interlopers. 

Our spectroscopic completeness is important for studying the
kinematics of galaxies with and without emission lines.  There is
little evidence for kinematic segregation between the two populations
within the virial radius $r_{200}$, contrary to previous results and
conventional wisdom.  We show that kinematic segregation is a very
subtle issue strongly dependent on membership classification, survey
depth and completeness, parametric versus nonparametric tests, and
galaxy classification.

Galaxy populations vary with environment both inside and outside
cluster virial regions.  Galaxies in cluster infall regions are an
intermediate population between cluster galaxies and field galaxies.
Thus, either the physical mechanisms responsible for changes in galaxy
populations operate outside of cluster virial regions or a significant
number of galaxies in the infall region have already passed through
the virial region.  The latter ``backsplash'' scenario has been
invoked to explain the presence of galaxies apparently stripped of H\I
~at large distances from clusters.  Numerical simulations predicts
different velocity distributions for backsplash galaxies and those on
first infall in the interval $1.4<R_p/r_{200}<2.8$ \citep{gill04}.  A
toy model of this backsplash scenario (where we assume that backsplash
is the only mechanism affecting star formation and that this mechanism
is perfectly efficient) disagrees with the observations.  The
H\I-deficient galaxies may in fact be backsplash galaxies, but our
result shows that the backsplash scenario cannot be the only or even
the primary mode of galaxy transformation.  This result supports
mechanisms for which local galaxy density is more important than the
distance to the nearest cluster.  This conclusion agrees with the
results of previous studies which show similar relations between
galaxy population and local density both near and far from clusters
\citep{lewis02,gomez03,balogh03,treu03,tanaka04}.  We confirm that
galaxy properties seem to change significantly at local densities
larger than $\Sigma _5\gtrsim 2 h^2 \mbox{Mpc}^{-2}$.  It is
remarkable that a similar scale has been found in many heterogeneous
studies using different wavelengths for selection and different
classification schemes (morphological, photometric, or spectroscopic)
\citep{1984ApJ...281...95P,lewis02,gomez03,balogh03,treu03,tanaka04}.

The {\it distribution} of star formation rates (emission-line
strengths in the inner parts of galaxies, to be precise) among
galaxies {\it with current star formation} shows {\it little}
dependence on environment, confirming previous studies
\citep{bcarter,balogh03,tanaka04}.  This result excludes mechanisms
which produce gradual reductions in star formation rates and supports
those that lead to sudden truncation of star formation \citep[see
also][]{vandokkum98,balogh03} or mechanisms (of any timescale) which
operate only until moderate redshifts \citep[timescales $\gtrsim$1.5
Gyr, see $\S \ref{distribha}$ and][]{kauffmann04}.  If these
mechanisms lose effectiveness in more recent times, star-forming
galaxies which ``survive'' until moderate redshift would evolve
similarly to star-forming field galaxies; galaxies that
lose their gas reservoir evolve until little evidence remains of
their earlier star formation.  An alternative possibility is that much
of the star formation seen in cluster virial and infall regions occurs
in bursts with short timescales consistent with tidal triggering
\citep[e.g.,][]{2000ApJ...530..660B,moss00}.  Finally, if most
emission-line galaxies observed at large projected densities lie at
significantly lower spatial densities, the similarity of the
distributions is simply a result of looking at two samples of the same
parent distribution.  From the number density profiles of the two
galaxy types, we calculate that at least 50\% of emission-line
galaxies with $R_p<r_{200}$ are infall interlopers.  Galaxies in
infall regions sample a wide range of local densities (Figure
\ref{s5r200}), so this explanation is plausible.

Our observations (and those of fiber-based surveys) are not very
sensitive to \ha emission in the galaxy outskirts; thus our spectra
will likely not show evidence of truncated star-forming disks
\citep[cf.][]{koopmann04b,vogt04b}.  If star formation in the inner
few kpcs of galaxies is unaffected on average, then this difference
might explain why we find a similar distribution of SFRs inside and
outside clusters.  Integrated spectra or wide-field, narrowband \ha
observations could test this hypothesis.  An objective prism study of
cluster galaxies suggests that cluster galaxies have a higher
incidence of tidally induced circumnuclear starbursts than field
galaxies \citep{moss00}, again indicating differences between the star
formation properties of cluster galaxies measured on different size
scales even though these differences are not large for field galaxies
\citep{kewley05}.

Future studies of galaxy properties in poor clusters and groups may
provide further insight into the physical mechanisms responsible for
the relation between star formation and local galaxy density.
Similarly, detailed studies of individual galaxies may show these
mechanisms in operation \citep[e.g.,][]{2004AJ....127.3361K}.
Extending these observations to systems at moderate and high redshift
would directly probe the timescale of this evolution.  In a study of a
cluster at $z\approx$0.4 with narrowband \ha~photometry,
\citet{kodama04} find trends similar to those found in the local
universe in both the spectral type-density relation and the weak
dependence of the \ha luminosity function on local density.  Finally,
H\I ~observations of complete samples of cluster galaxies could reveal
the connections between star formation rates and the available gas
reservoir
\citep[e.g.,][]{vogt04b}.  Untangling the relation between and
evolution of different tracers of star formation as a function of
environment is a difficult proposition, but future observations hold
the promise of resolving several of the curious results we find in
nearby clusters and their outskirts.

\acknowledgements

We thank Michael Balogh, Jeff Kenney, and Richard Larson for useful
discussions which significantly improved the presentation of this
paper.  We thank Perry Berlind and Michael Calkins, the remote
observers at FLWO, Susan Tokarz, who processed the spectroscopic data,
and all FAST queue observers who took data for CAIRNS.  MJG and MJK
are supported in part by the Smithsonian Institution.  This research
has made use of the NASA/IPAC Extragalactic Database (NED) which is
operated by the Jet Propulsion Laboratory, California Institute of
Technology, under contract with the National Aeronautics and Space
Administration.  This publication makes use of data products from the
Two Micron All Sky Survey, which is a joint project of the University
of Massachusetts and the Infrared Processing and Analysis
Center/California Institute of Technology, funded by the National
Aeronautics and Space Administration and the National Science
Foundation.  We thank the anonymous referee for a careful reading and
useful questions and suggestions which improved the presentation of
this paper.

\bibliographystyle{apj}
\bibliography{rines}

\clearpage
\clearpage
\clearpage
\clearpage
\clearpage
\clearpage
\clearpage
\clearpage
\clearpage
\clearpage
\clearpage
\clearpage
\clearpage
\clearpage
\clearpage
\clearpage
\clearpage
\clearpage

\end{document}